\documentclass[aps,prb,twocolumn,amsmath,amssymb,superscriptaddress,floatfix,showpacs]{revtex4}
\usepackage{times}
\usepackage{color}

\usepackage[dvips]{graphicx}
\begin{document}
\title{Nematic, vector--multipole, and plateau--liquid states in the classical $O(3)$ pyrochlore antiferromagnet
      with biquadratic interactions in applied magnetic field
}
\author{Nic Shannon}
\affiliation{H.\ H.\ Wills Physics Laboratory, University of Bristol,  Tyndall Av, BS8--1TL, UK.}
\author{Karlo Penc}
\affiliation{
Research Institute  for  Solid State  Physics   and
Optics, H--1525 Budapest, P.O.B.~49, Hungary}
\author{Yukitoshi Motome}
\affiliation{Department of Applied Physics, University of Tokyo, 
Bunkyo--ku, Tokyo 113--8656, Japan}
\date{\today}
\begin{abstract}   
The classical bilinear--biquadratic nearest--neighbor Heisenberg antiferromagnet on the pyrochlore lattice
does not exhibit conventional N\'eel--type magnetic order at any temperature or magnetic field.
Instead spin correlations decay algebraically over length scales $r \lesssim \xi_c \sim \sqrt{T}$, behavior
characteristic of a Coulomb phase arising from a strong local constraint.     
Despite this,  its thermodynamic properties remain largely 
unchanged if N\'eel order is restored by the addition of 
a degeneracy--lifting perturbation, e.g., further neighbor interactions.   
Here we show how these apparent contradictions can be resolved by a proper understanding of
way in which long--range N\'eel order emerges out of well--formed local correlations, and identify 
nematic and vector--multipole orders hidden in the different Coulomb phases of the model.
So far as experiment is concerned, our results suggest that where long range interactions are 
unimportant, the magnetic properties of Cr spinels which exhibit half--magnetization plateaux may 
be largely independent of the type of magnetic order present.
\end{abstract}
\maketitle

\section{Introduction}

Frustrated magnets have long been studied as a paradigm for complex 
behavior in condensed matter and statistical physics~\cite{physicstoday}.
The most widely studied systems are frustrated antiferromagnets (AF), where
competing interactions suppress classical N\'eel order.   In some
highly frustrated magnets, spins do not order at {\it any} temperature,
and the ground state retains only very short--ranged spin--spin correlations.
The resulting state is generally termed a ``spin liquid''.

However it is also possible to invert this paradigm and think of highly frustrated
magnets as systems where local ``order'' is robust enough to survive,
even where long range order has been obliterated by fluctuations.   
Conventional N\'eel order can then easily be restored --- 
albeit with a relatively low critical temperature 
---
by {\it any} perturbation which forces long--range coherence on this preformed local order. 
Moreover, where quantities such as heat capacity and magnetic 
susceptibility are controlled by local fluctuations, the thermodynamic properties of the globally--disordered
spin--liquid phase may be practically indistinguishable from those of the magnetically ordered phase.

\begin{figure}[t]
\includegraphics[height=5.5cm]{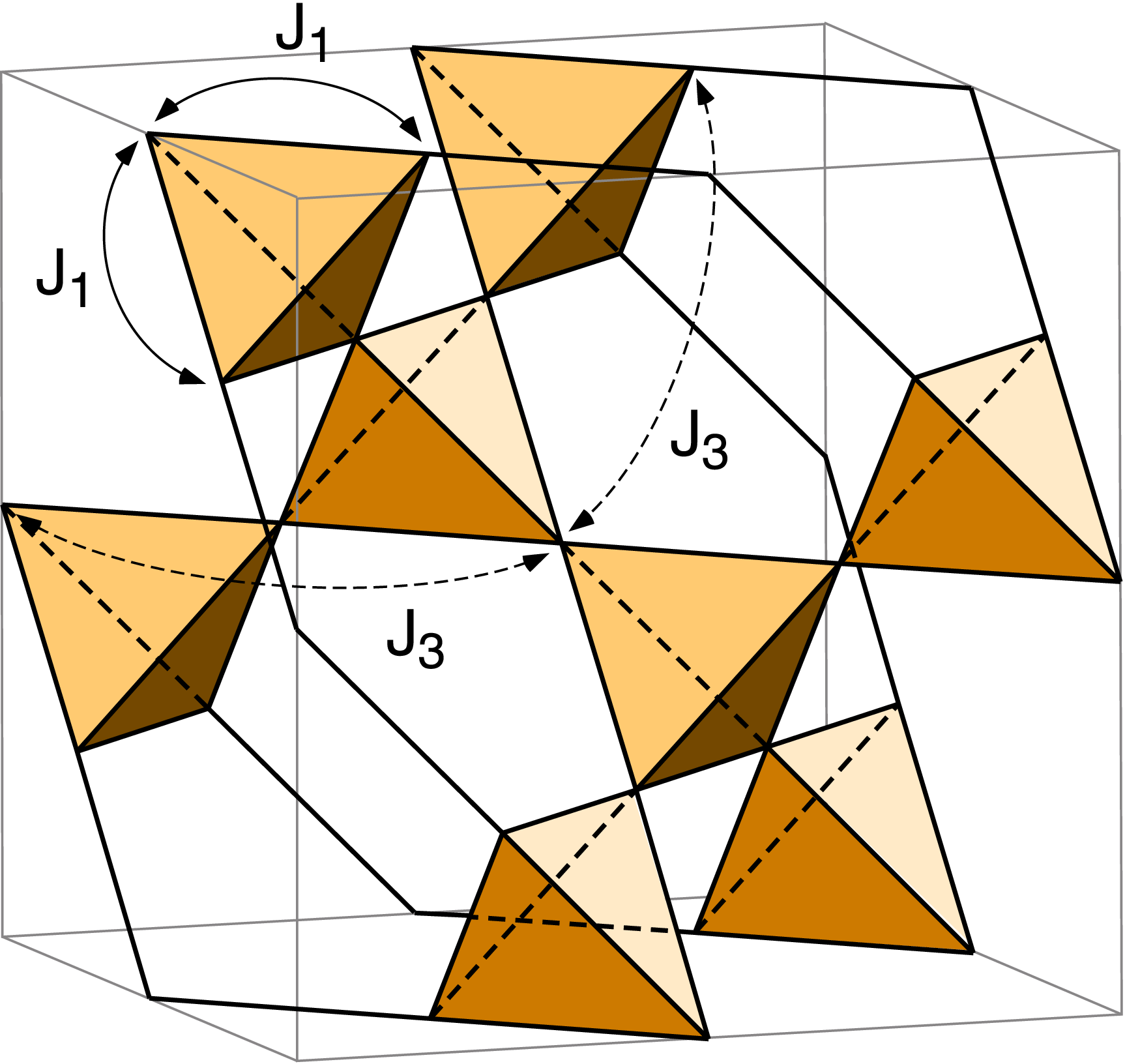}
\caption{(Color online) \label{fig:pyrochlore} 16-site cubic unit cell of the pyrochlore lattice --- a network
of corner--sharing tetrahedra.  Exchange interactions $J_1$ are associated with the first neighbor bonds $\langle ij 
\rangle_1$, and $J_3$ with the two (inequivalent) types of third--neighbor bond $\langle ij \rangle_3$.}
\end{figure}

In this paper we explore this ``bottom--up'' formulation of frustration, showing how the different multipolar and 
spin liquid states of a simple classical frustrated antiferromagnet in applied magnetic field
already contain the seeds of long range N\'eel order.   
The model which we consider  is the antiferromagnetic nearest--neighbor 
Heisenberg model with additional biquadratic interactions $b$ 
\begin{eqnarray}
    \label{eqn:H0}
     \mathcal{H} &=& J_1 \sum_{\langle ij \rangle_1} \big[ {\bf S}_i \cdot {\bf S}_j
       - b \, ({\bf S}_i \cdot {\bf S}_j)^2 \big]
	   -  {\bf h} \cdot \sum_i {\bf S}_i \, ,
	      \label{eq:Hb} 
\end{eqnarray}
where the sum $\langle ij \rangle_1$ runs over the nearest neighbor bonds 
of a pyrochlore lattice (Fig.~\ref{fig:pyrochlore}). All the energy scales including $h \equiv |{\bf h}|$ 
and temperature $T$ are measured in units of $J_1$ hereafter.

This model was introduced in Ref.~[\onlinecite{penc04}] to explain the dramatic 
half--magnetization plateau observed in Cr spinels~\cite{ueda05,ueda06,shannon07,matsuda07,ueda08,kojima08}.
In this case the biquadratic interaction $b$ originates in a strong 
coupling to the lattice.  However such terms can also be of electronic 
origin, and quite generally  they can be taken to
characterize the effects of quantum and/or thermal fluctuations in a frustrated 
magnet~\cite{henley87,larson09,nikuni93}. Thus we anticipate many of our results will also be relevant for the 
quantum model.  Recent results for the $S=3/2$ $XXZ$ pyrochlore AF in applied magnetic field 
suggest that this is indeed the case~\cite{bergman05,bergman06}.

\begin{figure}[t]
	\begin{center}
    \includegraphics[width=7.5truecm]{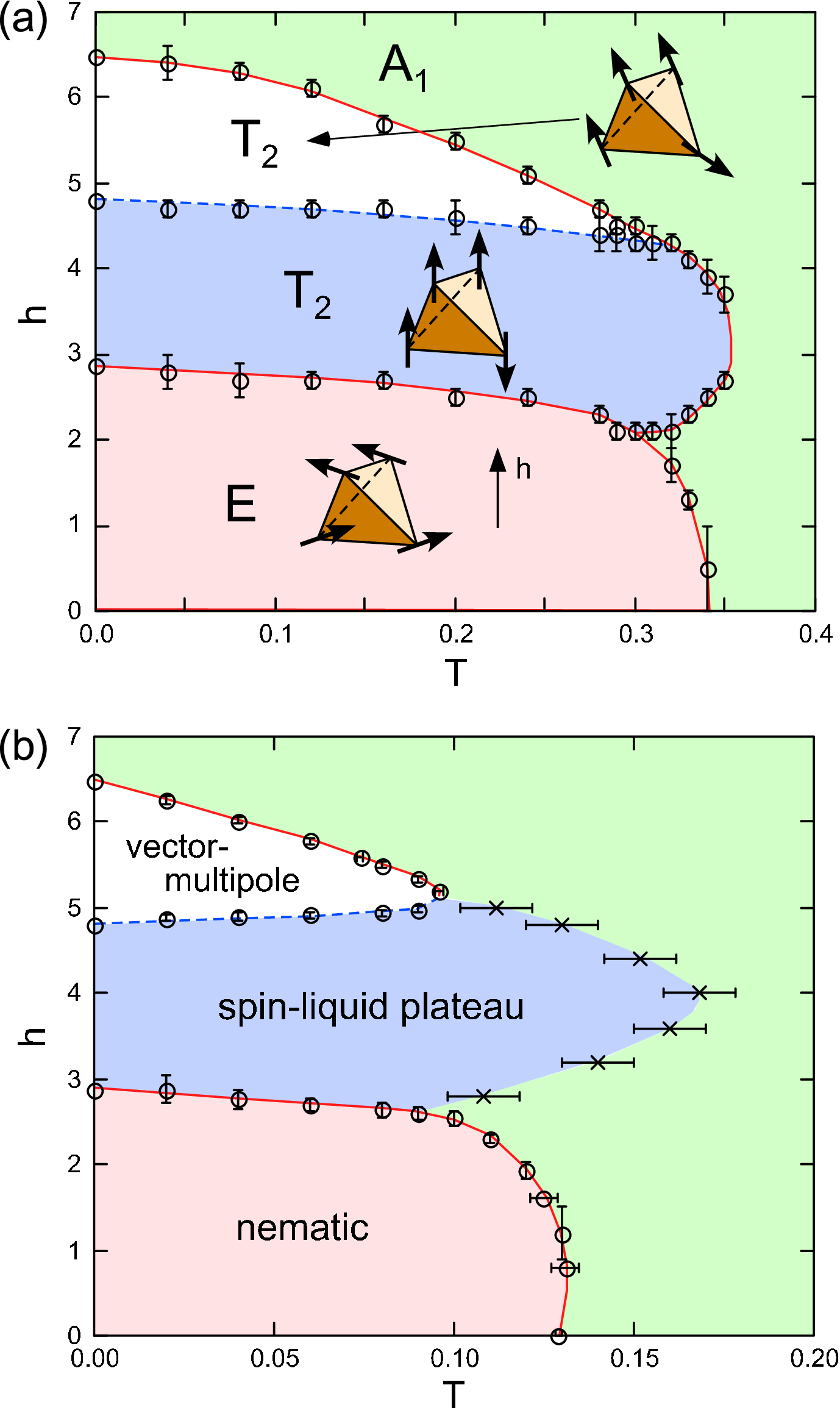}
	\end{center}
	\caption{(Color online)  
(a)~Magnetic phase diagram of the classical pyrochlore antiferromagnet
with biquadratic interactions 
$b=0.1$, and additional third--neighbor 
interactions \mbox{$J_3 = -0.05$}, 
as determined by classical Monte Carlo simulation~\protect\cite{motome05, motome06}.  
The form of four--sublattice N\'eel order is illustrated,  together with the irreducible representation 
(irrep) of the tetrahedral symmetry group ${\cal T}_d$ to which it belongs.   
For these parameters, the model provides a good description of the half-magnetization 
plateau seen in CdCr$_2$O$_4$~\protect\cite{ueda05,ueda06,shannon07,matsuda07,ueda08,kojima08}.
(b)~Equivalent magnetic phase diagram 
in the absence of any longer range interactions.  
Canted N\'eel states are replaced by phases with multipolar order,
while the collinear half-magnetization plateau state gives way to a collinear spin liquid.
Crosses denote the crossover at temperature $T^*$ 
from paramagnet to the plateau--liquid state, as 
determined by a peak in the heat capacity.
In both (a) and (b) circles with solid (red) lines denote first--order phase boundaries, 
while those with dashed (blue) lines denote second--order ones.
Both $T$ and $h$ are measured in units of $J_1$.
\label{fig:hTphasediag}}
\end{figure}

It is well known that the classical Heisenberg model 
with the nearest--neighbor bilinear couplings only 
does not exhibit N\'eel--type 
magnetic order on the pyrochlore lattice at any temperature~\cite{reimers92a,moessner98}. 
As we shall see, these arguments are essentially unchanged by the introduction of magnetic field,
or by nearest--neighbor biquadratic interactions $b$.
The system {\it can} however be brought to order by 
introducing an interaction which links spins in different tetrahedra, for example,
\begin{equation}
\label{eq:HJ3}
%\Delta 
{\mathcal H}^{\sf LRO} = J_3 \sum_{\langle ij \rangle_3} {\bf S}_i \cdot {\bf S}_j \, , 
\end{equation}
where $\langle ij \rangle_3$ runs over the two (inequivalent) sets of third neighbor 
bond shown in Fig.~\ref{fig:pyrochlore}.   

For ferromagnetic (FM) $J_3 < 0$, this specific form of 
${\mathcal H}^{\sf LRO}$ leads to the four--sublattice 
long-range order (LRO) described in Ref.~[\onlinecite{penc04}], 
and to the finite temperature transitions shown in Fig.~\ref{fig:hTphasediag}(a)~\cite{motome05, motome06}.     
Four sublattice order can also be stabilized by AF second neighbor interaction 
$J_2$.  More generally, however, the type of order which results depends on the details 
of the interaction 
${\mathcal H}^{\sf LRO}$~\cite{reimers1991,yaresko08}.   
The system can therefore be tuned at will 
between different types of ordered state, simply by changing 
${\mathcal H}^{\sf LRO}$.
From this we conclude that, as a function of magnetic field $h$, for 
${\mathcal H}^{\sf LRO}= 0$, 
there must be a line of second--order {\it multicritical} --- or first--order {\it multifurcative} points --- 
separating a huge set of different ordered phases. 

The main purpose of this paper is to explore the symmetry breaking  which {\it persists} in the limit of
${\mathcal H}^{\sf LRO} \to 0$ for finite biquadratic interaction $b$ and  finite temperature $T$.  
In order to make the problem accessible to large scale Monte Carlo 
(MC) simulation, we consider the classical $S \equiv |{\bf S}| \to \infty$ limit of 
Eq.~(\ref{eq:Hb}), rescaling variables such that $S \equiv 1$. 

Using a mixture of classical MC simulation, analytic low--$T$ expansion, and 
simple field theoretical arguments, we find a set of phases in the $h$--$T$ plane which exhibit 
power--law decay of spin correlation functions.  
Two of these phases possess long--range nematic or vector--multipole order and, most interestingly,
the magnetization plateau persists in the absence of conventional magnetic order.   
We show how all of these results can be understood --- and even anticipated ---
from a proper understanding of the geometry of the pyrochlore lattice, and the way in which a single tetrahedron 
behaves in magnetic field.  Our findings are summarized by the $h$--$T$ phase diagram 
shown Fig.~\ref{fig:hTphasediag}(b).

So far as experiment is concerned, our main conclusion will be that 
the thermodynamic properties of the pyrochlore antiferromagnet
in applied magnetic field are mostly determined by symmetry breaking at the level of single tetrahedron.   
Local order is well--formed 
for ${\mathcal H}^{\sf LRO}= 0$, and many properties 
of the system are therefore insensitive to the details of the LRO order present.   
Thus the very simple phase diagram derived in Ref.~[\onlinecite{penc04}] and its finite 
temperature generalization in Refs.~[\onlinecite{motome05}] and [\onlinecite{motome06}]
[reproduced in Fig.~\ref{fig:hTphasediag}(a)], are applicable for a wide variety of different 
${\mathcal H}^{\sf LRO}$.

The paper is structured as follows:  In Sec.~\ref{degeneracies} we briefly review the basic
physics of the Heisenberg model on the pyrochlore lattice.  Definitions are given of order parameters for 
conventional N\'eel (dipolar) order, and of rank--two tensor order parameters which can be 
used to signal multipolar order.
\begin{figure*}
    \centering
    \hfill
    \includegraphics[width=.4\textwidth]{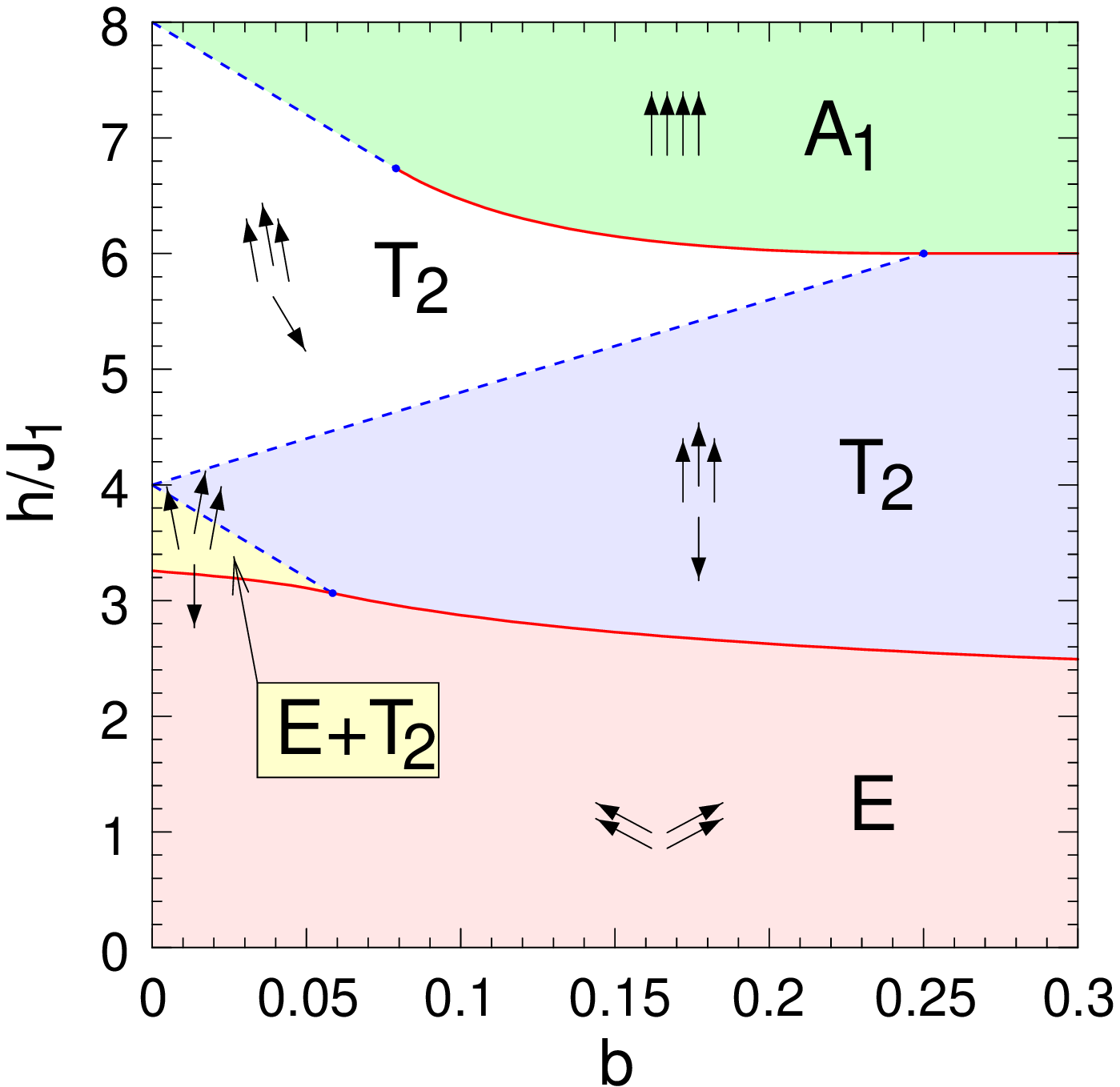}
    \hfill
    \includegraphics[width=.4\textwidth]{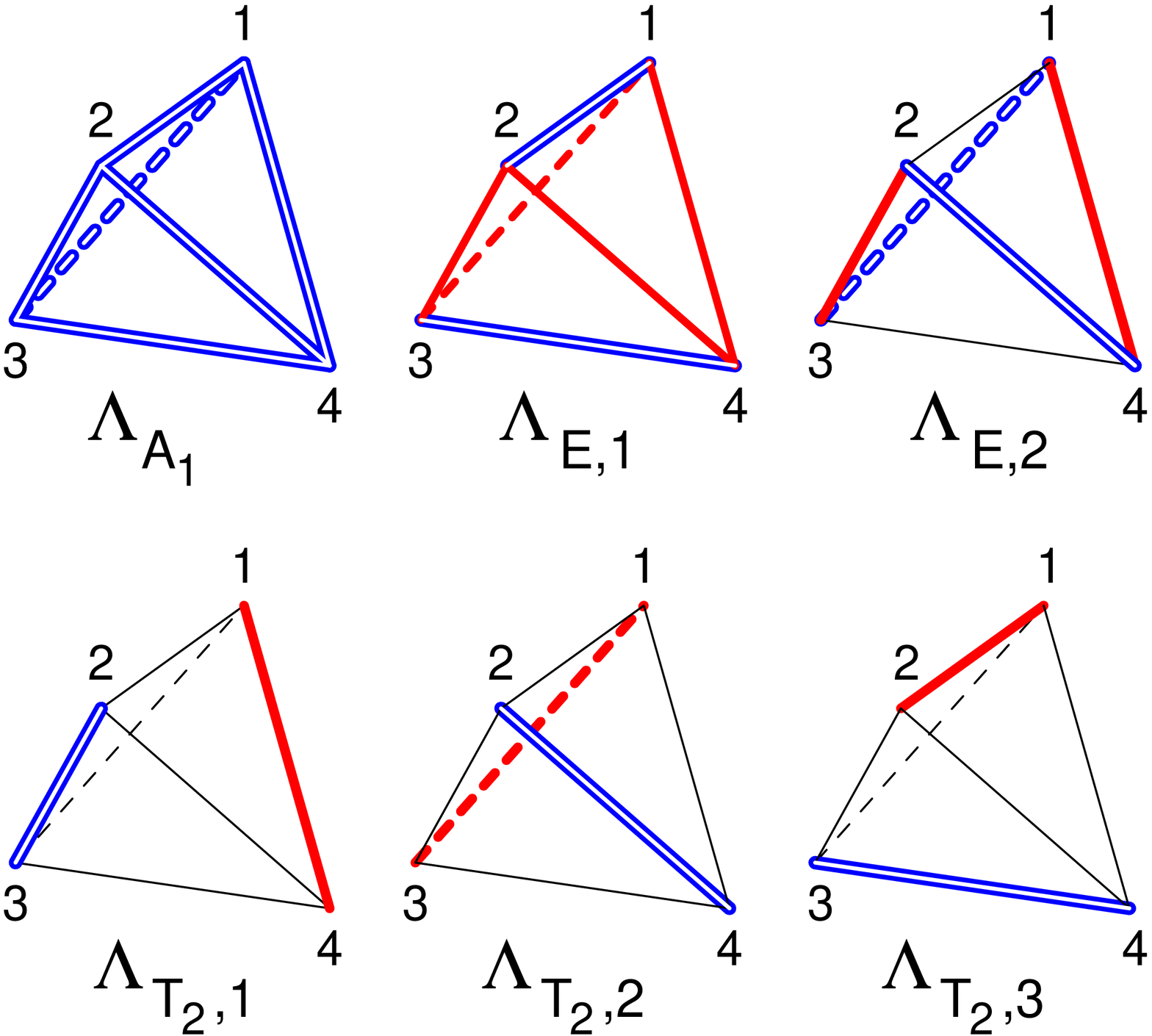}
    \hfill\null \caption{(Color online)  
 (Left panel)~Gound--state phase diagram of a single classical tetrahedron as a function of
  magnetic field $h$ and dimensionless coupling constant $b$, 
  taken from Ref.~[\onlinecite{penc04}].  Solid (red) lines denote first and
  dashed (blue) lines second order transitions. Spin configurations and relevant 
  irreducible representations (irreps) are shown in each case. 
(Right panel)~Symmetries of the ${\sf A_1}$, ${\sf E}$, and ${\sf T_2}$ irreps of 
  the tetrahedral group ${\cal T}_d$ used to classify different states.  
  Solid (red) lines have negative weight; 
  hollow (blue) lines have positive weight.   Thin (black) lines have zero weight.  
  See also Eq.~(\ref{eq:Tdsym}). 
}
    \label{fig:tetrahedron}
\end{figure*}

Then, in Sec.~\ref{deltaJzero} we use these tools to construct 
the \protect\mbox{$h$--$T$} phase diagram of the pyrochlore AF with additional biquadratic interactions [Fig.~\ref{fig:hTphasediag}(b)].    
Thermal fluctuations preserve the extensive degeneracies present in the ground state, 
and fail to select any conventional long--range dipolar order.  Despite this, the thermodynamic properties 
of the system and the topology of the phase diagram are essentially unchanged
--- the magnetization plateau survives and nematic and vector--multipole phases corresponding to the 
two different canted states are shown to exist for fields below and above the magnetization plateau
[Fig.~\ref{fig:hTphasediag}(a)].

In Sec.~\ref{deltaJfinite} we explore the way in which long--range
N\'eel  order is recovered as a FM third--neighbor
interaction $J_3$ is ``turned on'', focusing on the half--magnetization plateau
for $h \approx 4$.   
For small $|J_3|$, the system now exhibits two characteristic
temperature scales --- an upper temperature $T^* \approx b$ at which the gap 
protecting the magnetization plateau opens, and a lower temperature 
$T_N \approx {\cal O}(|J_3|)$ at
which the system exhibits long--range magnetic order.   
This is contrasted with the situation for $h=0$, 
where the system also exhibits two characteristic temperature scales, 
but these correspond to successive phase transitions :
a nematic transition at $T_Q \sim b$ and a N\'eel ordering at $T_N \sim O(|J_3|)$.
We discuss the nature of these transitions for $J_3 \to 0$, identifying a line of first--order 
{\it multifurcative} points at $J_3=0$.    And, for $J_3=0$, we identify 
an unusual continuous transition from the coulombic plateau liquid to the vector-multipole phase. 
At low temperatures this transition appears to have mean field character.  

Finally, in Sec.~\ref{conclusions} we conclude with a discussion of the broader implications 
of these results.

%%%%%%%%%%%%%%%%%%%%%%%%%%%%%%%%%%%%%%%%%%%%

\section{Degeneracies in finite magnetic field}
\label{degeneracies}

%%%%%%%%%%%%%%%%%%%%%%%%%%%%%%%%%%%%%%%%%%%%

\subsection{Geometrical arguments}

The pyrochlore lattice~(Fig.~\ref{fig:pyrochlore}) is the simplest example of a three--dimensional (3D)
network of corner sharing complete graphs.   Its elementary building block is the tetrahedron, 
in which every site is connected to every other site, i.e. the tetrahedron is a complete graph of order four.   
Tetrahedra in the pyrochlore lattice can be divided into A and B sublattices, with each lattice site shared 
between an A- and a B-sublattice tetrahedron.   The centres of the two types of tetrahedra together form 
a (bipartite) diamond lattice.~\footnote{The pyrochlore lattice can also be thought of as bi-simplex, where is each tetrahedron 
is a simplex.}  The overall symmetry of the lattice is cubic.

As such, the pyrochlore lattice is a natural 3D analogue of the 2D kagome lattice, a corner sharing network
of triangles (complete graphs of order three).   In fact the [111] planes of the pyrochlore lattice are alternate
kagome and triangular lattices, composed of the triangular ``bases'' of tetrahedra and their ``points'', respectively.
Much of the unusual physics of the kagome lattice also extends to its higher dimensional cousin.

Lattices composed of complete graphs have the special property that bilinear quantities on nearest neighbor
bonds can be recast as a sum of squares.   Thus for  $b=0$ the Hamiltonian Eq.~(\ref{eqn:H0}) can be written
\begin{equation}
\label{eqn:Hsquare}
 \mathcal{H} = 4 \sum_{\sf tetra} 
  \left({\bf M} - \frac{{\bf h}}{8}\right)^2 -\frac{h^2}{16} 
  + {\text{const.}} \, ,
\end{equation}
where the sum runs over tetrahedra, and 
\begin{equation}
 {\bf M} = \frac{1}{4}({\bf S}_1 + {\bf S}_2 + {\bf S}_3 +  {\bf S}_4)
\end{equation}
is the magnetization (per site) of a given tetrahedron.   
For \mbox{$h=0$}, a simple classical counting argument shows that two of the eight 
angles needed to determine the orientation of the four spins in any given tetrahedron 
remain undetermined.    
Nearest neighbor interactions do not select one unique ground state on the pyrochlore lattice 
but rather the entire {\it manifold} of states for which  $|{\bf M}| = 0$ in {\it each} tetrahedron.
Thus at $T=0$, the system is disordered.
For fields $h < h_{\rm sat} = 8$ this conclusion is unaltered by the presence of magnetic field.
In this case the manifold of ground states is  determined by the condition ${\bf M} = {\bf h}/8$ 
in each tetrahedron, and the magnetization is linear in $h$ up to 
the saturation field $h_{\rm sat}=8$.  (We recall that magnetic field is measured in units of $J_1$, 
so that in fact $h_{\rm sat}=8 J_1$.)

In order to understand how nearest--neighbor biquadratic interactions $b$ select among this 
manifold of states, it is sufficient to solve the problem of a single tetrahedron embedded
in the 3D lattice.   This problem was considered in Ref.~[\onlinecite{penc04}].  
For \mbox{$b>0$}, biquadratic interactions select coplanar (and collinear) configurations from 
the larger ground--state manifold of Eq.~(\ref{eqn:Hsquare}).  There are four dominant phases, 
illustrated in Fig.~\ref{fig:tetrahedron}~:
\begin{enumerate}
\item[(i)] a 2:2 coplanar canted state for low field 
\item[(ii)]  a 3:1 collinear ($uuud$) half--magnetization plateau state for intermediate field 
\item[(iii)]  a 3:1 coplanar canted state for fields approaching saturation
\item[(iv)]  a saturated ($uuuu$) state for large magnetic field \mbox{$h>h_{\rm sat}$}
\end{enumerate}
An exhaustive enumeration of possible states is given in Ref.~[\onlinecite{penc07}].

Up to this point, we have not been specific about {\it how} the tetrahedron was embedded in the lattice.
It could, trivially, form part of a state with N\'eel order, e.g., the simple four--sublattice order favored by
FM $J_3$.   However, there are infinitely many other ways of joining 2:2 or 3:1 tetrahedra together
at the corners, and not all of them correspond to N\'eel ordered states.   In fact the ground state
manifold retains an extensive Ising--like degeneracy for all $h<h_{\rm sat}$, and as a result the system 
remains ``disordered''.   The nature of this degeneracy, and its consequences, are explored in some detail below.

%%%%%%%%%%%%%%%%%%%%%%%%%%%%%%%%%%%%%%%%%%%%

\subsection{Bond order parameters}
\label{orderparameters}

Where N\'eel order is present, it can be detected in the reduced spin--spin correlation function
\begin{equation}
    D({\bf r}_{ij}) = \langle {\bf S}_i \cdot {\bf S}_j \rangle 
      - m^2 \, .
    \label{eq:spin-spin}
\end{equation}
Here $m^2$ is the expectation value of the squared magnetization per spin,
\begin{equation}
m^2 = \Big\langle \frac{1}{N}  \big( \sum_i {\bf S}_i \big)^2 \Big\rangle \, ,
\label{eq:m}
\end{equation} 
which vanishes in the absence of magnetic field.   $N$ is the total number of spins.  
The simplest form of order supported by the pyrochlore lattice is the four--sublattice N\'eel order
favored by FM $J_3$, as illustrated in Fig.~\ref{fig:hTphasediag}(a).  

Written in terms of the minimal four--site unit cell of the pyrochlore lattice, four--sublattice order has 
momentum \protect\mbox{${\bf q} =0$}, and different states can easily be classified using the 
${\sf A_1}$, ${\sf E}$, and ${\sf T_2}$ irreducible representations (irreps) of the 
symmetry group ${\cal T}_d$ for a single tetrahedron~:
\begin{widetext}
\begin{equation}
 \left(
 \begin{array}{l}
  \Lambda_{{\sf {\sf A_1}}}  \\
  \Lambda_{{\sf E},1}  \\
  \Lambda_{{\sf E},2}  \\
  \Lambda_{{\sf T_2},1}  \\
  \Lambda_{{\sf T_2},2}  \\
  \Lambda_{{\sf T_2},3}  \\
 \end{array}
\right)=
\left(
\begin{array}{cccccc}
\frac{1}{{\sqrt{6}}} & \frac{1}{{\sqrt{6}}} & \frac{1}{{\sqrt{6}}} &
 \frac{1}{{\sqrt{6}}} & \frac{1}{{\sqrt{6}}} & \frac{1}{{\sqrt{6}}} \\
\frac{1}{\sqrt{3}} & \frac{-1}{2\sqrt{3}} & \frac{-1}{2\sqrt{3}} &
 \frac{-1}{2\sqrt{3}} & \frac{-1}{2 \sqrt{3}} & \frac{1}{\sqrt{3}} \\
0 & \frac{1}{2} & -\frac{1}{2}  & -\frac{1}{2}  & \frac{1}{2} & 0 \\
0 & 0 & \frac{-1}{\sqrt{2}} & \frac{1}{\sqrt{2}} & 0 & 0 \\
0 & \frac{-1}{\sqrt{2}} & 0 & 0 & \frac{1}{\sqrt{2}}  & 0 \\
\frac{-1}{\sqrt{2}} & 0 & 0 & 0 & 0 & \frac{1}{\sqrt{2}} \\
\end{array}
\right)
\left(\begin{array}{c} 
{\bf S}_1 \cdot {\bf S}_2 \\
{\bf S}_1 \cdot  {\bf S}_3 \\
{\bf S}_1 \cdot {\bf S}_4 \\ 
{\bf S}_2 \cdot {\bf S}_3\\
{\bf S}_2 \cdot {\bf S}_4\\
{\bf S}_3 \cdot {\bf S}_4\\
\end{array}
\right) \, , 
\label{eq:Tdsym}
\end{equation}
\end{widetext}
where the spins ${\bf S}_i$ belong to a single tetrahedron~\cite{penc04,penc07,tchernyshyov02}  
--- cf. right panel of Fig.~\ref{fig:tetrahedron}.
We can use these irreps to define bond order parameters 
\begin{eqnarray}
\lambda_\nu^{\sf global} &=& \frac{4}{N}
\Big( \,
\sum_{\sf tetra} {\bf \Lambda}_{\nu}
\, \Big)^2 \, ,
\label{eq:global}
\end{eqnarray}
and associated generalized susceptibilities
\begin{eqnarray}
 \chi_\nu^{\sf global} &=& 
\frac{N}{T} \big[ \langle (\lambda_\nu^{\sf global})^2 \rangle 
- \langle \lambda_\nu^{\sf global} \rangle^2 \big] \, ,
 \label{eq:global_chi}
\end{eqnarray}
where the sum $\sum_{\sf tetra}$ runs over all 
$N/4$ independent A--sublattice tetrahedra, 
and ${\bf \Lambda}_{\nu}$ is the vector associated 
with the \mbox{$\nu = \{{\sf E}, {\sf T_2}\}$} irreps of the tetrahedral 
symmetry group~${\cal T}_d$, namely,
${\bf \Lambda}_{\sf E} = (\Lambda_{\sf E,1},\Lambda_{\sf E,2})$ and
${\bf \Lambda}_{\sf T_2} = (\Lambda_{\sf T_2,1},\Lambda_{\sf T_2,2},\Lambda_{\sf T_2,3})$.  

These order parameters allow us to distinguish the sharp first--order 
transition between orders in the ${\sf T_2}$ and ${\sf E}$ irreps, and 
the transition from the ${\sf A_1}$ to ${\sf T_2}$ states at high 
field --- cf. Fig.~\ref{fig:hTphasediag}(a) --- but not the more subtle second--order transition between the ${\sf T_2}$ symmetry 
$uuud$ plateau state and the ${\sf T_2}$ symmetry  3:1 canted state.
These are none the less distinct phases --- the collinear and canted ${\sf T_2}$ states
are connected by a zone--center (i.e., ${\bf q} = {\bf 0}$) excitation which is 
gapped in the collinear $uuud$ state, and becomes soft at the critical field 
marking the onset of the 3:1 canted state~\cite{penc07}. 
The condensation of this soft spin mode corresponds to the emergence 
of order in the transverse spin components $\langle S^x \rangle$, 
$\langle S^y \rangle$ --- i.e., canting of spins away from the $z$ axis
(the direction of applied magnetic field).

The bond order parameters defined in Eq.~(\ref{eq:Tdsym}) couple directly 
to the lattice, through changes in bond length~\cite{penc04}.   They are 
therefore particularly well suited to describing simple N\'eel ordered states, 
where the magnetic ordering is driven by the lattice effects.   However the
irreps on which they are based also provide a useful measure of the correlation 
which survives in the absence of long range order, a central question for this paper.
To this end, we introduce a measure of local correlation
\begin{equation}
 \lambda_\nu^{\sf local} = \frac{4}{N}\sum_{\sf tetra} {\bf \Lambda}_{\nu}^2 \, ,
 \label{eq:local}
\end{equation}
 and its associated generalized susceptibility 
\begin{equation}
 \chi_\nu^{\sf local} = 
\frac{N}{T} \big[ \langle (\lambda_\nu^{\sf local})^2 \rangle 
 - \langle \lambda_\nu^{\sf local} \rangle^2 \big] \, .
 \label{eq:local_chi}
\end{equation}
For a single tetrahedron, $\lambda_\nu^{\sf local}$ and $\lambda_\nu^{\sf 
global}$ are identical.   On a lattice, $\lambda_\nu^{\sf local}$ 
lacks crossterms between different tetrahedra present in $\lambda_\nu^{\sf global}$, and 
is therefore a measure of correlation in the absence of long range order.  
We return to these points below.

\subsection{Rank--two tensor order parameters}
\label{sec:nematic_order_parameter}

Not all of the phases supported by the Hamiltonian Eq.~(\ref{eq:Hb}) can be described using the bond order parameters 
Eq.~(\ref{eq:Tdsym}).  In Appendix~\ref{symmetry} we formally classify the different types of symmetry breaking which can 
arise in this model at the level of a single site.     
Here we restrict ourselves to the simplest possible generalization from N\'eel to multipolar order; 
both the ${\sf T_2}$ and ${\sf E}$ 
symmetry canted states possess order of transverse (i.e., $x$ and $y$)
spin components which vanishes in the collinear $uuud$ state, and which can 
survive even in the absence of conventional (canted) N\'eel order.  

To describe this, it is convenient to introduce the rank--two tensor order parameters 
\begin{eqnarray}
Q^\alpha = \frac{1}{N} \sum_{i=1}^N Q^\alpha_i \, ,
\end{eqnarray} 
where the local quadrupole moments
\begin{eqnarray}
Q^{3z^2-r^2}_i   &=&   \frac{1}{\sqrt{3}} \left[ 2(S^z_i)^2 - (S^x_i)^2 - (S^y_i)^2  \right]\, , \\
Q^{x^2-y^2}_i  &=&   (S_i^x)^2   -  (S_i^y)^2  \, ,  \\
Q^{xy}_i  &=& 2  S_i^x S_i^y  \, , \\
Q^{xz}_i  &=&   2  S_i^x S_i^z  \, , \\
Q^{yz}_i  &=&   2  S_i^y S_i^z \, , 
\label{eqn:quadrupoles}
\end{eqnarray}    
are summed over all lattice sites $i$.  

Where spin rotational symmetry is not already broken by magnetic field, i.e., for $h=0$, 
spins may select a common axis without selecting a direction on it.    This is conventional 
nematic order, of the type exhibited by uniaxial molecules, and can be detected using
the order parameter
\begin{eqnarray}
\label{eq:Q}
Q^2 &=& 
(Q^{3z^2-r^2})^2 + (Q^{x^2-y^2})^2 \nonumber\\
&& + \, (Q^{xy})^2+ (Q^{xz})^2 +(Q^{yz})^2 
\end{eqnarray}
which is invariant under $O(3)$ rotations.  
This order parameter takes on its maximal value $\langle Q^2 \rangle \rightarrow 4/3$ in a perfectly collinear state, 
such as the 2:2 state for $T\rightarrow 0$.    

In what follows we will also make use of the correlation function measuring collinearity
\begin{equation}
 P({\bf r}_{ij}) =  \frac{3}{2}\left[ \left({\bf S}_i \cdot {\bf S}_j \right)^2 -\frac{1}{3} \right] \, ,
    \label{eq:collineaity}
\end{equation}
considered in Ref.~[\onlinecite{moessner98}].   
As defined, \protect\mbox{$-1/2 \leq \langle P({\bf r}_{ij}) \rangle \leq1$}, taking on the value 
$\langle P({\bf r}_{ij}) \rangle = 0$ for uncorrelated spins.   In fact $P({\bf r}_{ij})$  can also be expressed 
in terms of quadrupolar operators as
\begin{eqnarray}
 P({\bf r}_{ij}) &=&  \frac{3}{4} \sum_\alpha Q^\alpha_i  Q^\alpha_j \, ,
    \label{eq:collineaity2}
\end{eqnarray}
and it follows that 
\begin{eqnarray}
Q^2 &=& \frac{1}{N^2} \sum_{ij}  \frac{4}{3}P({\bf r}_{ij}) \, .
\end{eqnarray}

At finite $h$, the $O(3)$ invariant correlation function Eq.~(\ref{eq:collineaity}) still provides 
a useful measure of collinearity, but does not by itself signal a broken symmetry.
In this case it is convenient to group quadrupoles according to way in which they transform 
under the remaining $O(2)$ rotations about the direction of magnetic field --- conventionally the $z$ axis.
We therefore consider 
\begin{eqnarray}
{\bf Q}^{\perp,2} &=& \big\{Q^{x^2-y^2},Q^{xy} \big\} \, ,
\label{eq:Qperp2} \\
{\bf Q}^{\perp,1}&=& \big\{ Q^{xz},Q^{yz} \big\} \, ,
\label{eq:Qperp1} \\
Q^{\perp,0} &=& Q^{3z^2-r^2} \, ,
\label{eq:Qperp0}
\end{eqnarray}
where the magnetic field is assumed to be parallel to the $z$ axis.  
Each of the separate irreps ${\bf Q}^{\perp,n}$ transforms like $\{\cos n\phi,\sin n\phi\}$ 
--- or equivalently, $e^{in\phi}$ --- where $\phi$ is the polar angle in the plane 
perpendicular to the magnetic field.    
They can therefore be used as order parameters to detect the $n$--fold 
breaking of rotational symmetry in the $xy$ plane.  
The conventional nematic order parameter with full $O(3)$ symmetry, Eq.~(\ref{eq:Q}), 
is given by the sum of squares
\begin{eqnarray}
\label{eq:Qsq}
Q^2 &=& (Q^{\perp,0})^2 + |{\bf Q}^{\perp,1}|^2 + |{\bf Q}^{\perp,2}|^2\, . \label{eq:Q2}
\end{eqnarray}
%In what follows, we write 
%\begin{eqnarray}
%Q^{\perp,1} = |{\bf Q}^{\perp,1}| \; &,& \;
%Q^{\perp,2} = |{\bf Q}^{\perp,2}|
%\end{eqnarray}

%%%%%%%%%%%%%%%%       TABLE           %%%%%%%%%%%%%%%%%%%

\begin{table} 
\caption{\label{tab:quadrupoles} Classification of tensor order operators according to rotational symmetry about a $z$ axis defined 
by magnetic field:
Each forms an irrep transforming like $e^{i n \phi}$, where $n$ is an integer and $\phi$ is the polar angle in the $xy$ plane.  
Also indicated are the finite values of the order parameters in the 2:2 and 3:1 canted states.} 
\begin{ruledtabular} 
\begin{tabular}{cccc} 
order par. &tensor operators& 2:2 & 1:3  \\ 
\hline
\hline
$e^{2i\phi}$    & $\{Q^{x^2-y^2},Q^{xy}\}$ & finite & finite \\ 
\hline
$e^{i\phi}$       & $\{Q^{xz},Q^{yz}\}$ & 0 & finite   \\ 
                    & $\{S^{x},S^{y}\}$ & 0 & 0   \\ 
\hline
1                 & $Q^{3z^2-r^2}$ & finite & finite   \\ 
                  & $S^z$ & finite & finite \\ 
\end{tabular} 
\end{ruledtabular} 
\end{table} 

In finite magnetic field, the one--dimensional irrep $Q^{\perp,0}$ does not contain any information about broken symmetries 
and can generally be discarded.   However the two--dimensional irreps ${\bf Q}^{\perp,1}$ and ${\bf Q}^{\perp,2}$ distinguish different ordered phases.  In the 2:2 canted phase the mean square value of 
${\bf Q}^{\perp,2}$ takes on a finite value 
\begin{eqnarray}
\langle |{\bf Q}^{\perp,2}|^2 \rangle &=& \langle (Q^{x^2-y^2})^2 + (Q^{xy})^2 \rangle > 0 
\, .
\end{eqnarray}
This is another form of nematic order of the transverse spin moments 
--- one transforming like  $e^{i2\phi}$ --- and 
reflects the fact that spins select a common plane in which to cant.
At the same time mean square value of ${\bf Q}^{\perp,1}$ --- 
which transforms as $e^{i\phi}$, i.e., a vector in the $xy$ plane --- vanishes.
  
Similarly, $\langle |{\bf Q}^{\perp,2}|^2 \rangle$ takes on a finite value in the 3:1 canted phase.   
However in this case the 3:1 asymmetry of the canted spin configuration defines a direction in the $xy$ plane,
and 
\begin{eqnarray}
\langle |{\bf Q}^{\perp,1}|^2 \rangle &=& \langle (Q^{xz})^2 + (Q^{yz})^2 \rangle > 0 
\,
\end{eqnarray}
is also finite.   
The 3:1 canted phase therefore possess a form of vector-multipole order.
These facts are summarized in Table~\ref{tab:quadrupoles}.  

In what follows we concentrate almost exclusively on phases which do not exhibit conventional
magnetic order, as defined by $D({\bf r})$ in Eq.~(\ref{eq:spin-spin}), and characterize these states 
using the rank--two tensor order parameters listed in Table~\ref{tab:quadrupoles}.   
%Since ${\bf Q}^{\perp,1}$ has a lower symmetry than ${\bf Q}^{\perp,2}$, we will consider it 
%to be the primary order parameter where both occur together.  
For further details of conventional N\'eel phases, and comparison with experiment, we refer the interested reader to Ref.~[\onlinecite{motome06}].    
Rank--three tensors which also occur as order parameters in the 
present model are discussed in Appendix~\ref{sec:octupoles}.

%%%%%%%%%%%%%%%%%%%%%%%%%%%%%%%%%%%%%%%%%%%%

\subsection{General considerations}

Many frustrated systems with disordered ground states manage none the 
less to order at finite temperature.   This effect is known as ``order from disorder'' 
and occurs where there is a net entropy gain in selecting one particular state 
out of the disordered manifold.   Entropy is gained where a given
spin configuration (typically, collinear or coplanar) has a higher density 
of low--energy excitations than its peers.   However this entropy gain must 
be sufficient to offset the entropy lost by choosing one state out of the 
manifold.   Where the ground state manifold has an extensive degeneracy,
this is a very strong constraint.  Order--from--disorder effects are known to select one particular 
N\'eel ordered ground state in e.g., the frustrated square lattice~\cite{weber03}, but fail to do so in the case 
of the more frustrated kagome lattice~\cite{zhitomirsky02,zhitomirsky08}.

Even where fluctuations fail to stabilize one particular N\'eel ground state, they can still
select a subset of states from the ground state manifold with a smaller --- but none the less
extensive --- degeneracy.   This subset (submanifold) of states will not exhibit the long range 
spin--spin correlations which are the hallmark of conventional N\'eel--type magnetic order.
However this does not necessarily mean that the system is truly disordered --- it may well
exhibit long range order of a more complex type.  

A good example of this second type of order--from--disorder
effect is provided by the nearest--neighbor classical $XY$ model on the pyrochlore lattice, where
thermal fluctuations lead to nematic order with broken spin--rotational symmetry, but 
power--law decay of spin--spin correlations~\cite{moessner98}.

In what follows we use the order parameters defined in Sec.~\ref{sec:nematic_order_parameter} to 
identify phases of Eq.~(\ref{eq:Hb}) which exhibit nematic order in the absence of N\'eel order.    
We focus chiefly on different forms of unconventional order found in magnetic field.   
Closely related studies in magnetic field have been made of 
the classical Heisenberg model on a kagome lattice~\cite{zhitomirsky02}, 
and classical $XY$ model on a checkerboard lattice~\cite{canals04}.   
In both these cases unconventional order is stabilized by thermal fluctuations.  Another type of unconventional order 
for the pyrochlore lattice with FM second--neighbor interactions $J_2 <0$ and $h=0$ was recently studied in Ref.~[\onlinecite{chern08}].   
In our case the main driving force is not fluctuations but finite biquadratic interaction $b$;   
results for order stabilized by thermal fluctuations at finite $h$ and $J_3$ but $b\equiv0$ 
will be presented elsewhere~\cite{motome09}.

%%%%%%%%%%%%%%%%%%%%%%%%%%%%%%%%%%%%%%%%%%%%

\section{Partial lifting of degeneracy in finite magnetic field}
\label{deltaJzero}

%%%%%%%%%%%%%%%%%%%%%%%%%%%%%%%%%%%%%%%%%%%%

%\begin{figure}[tb]
\begin{figure}[t]
  \centering
  \includegraphics[width=5.5truecm]{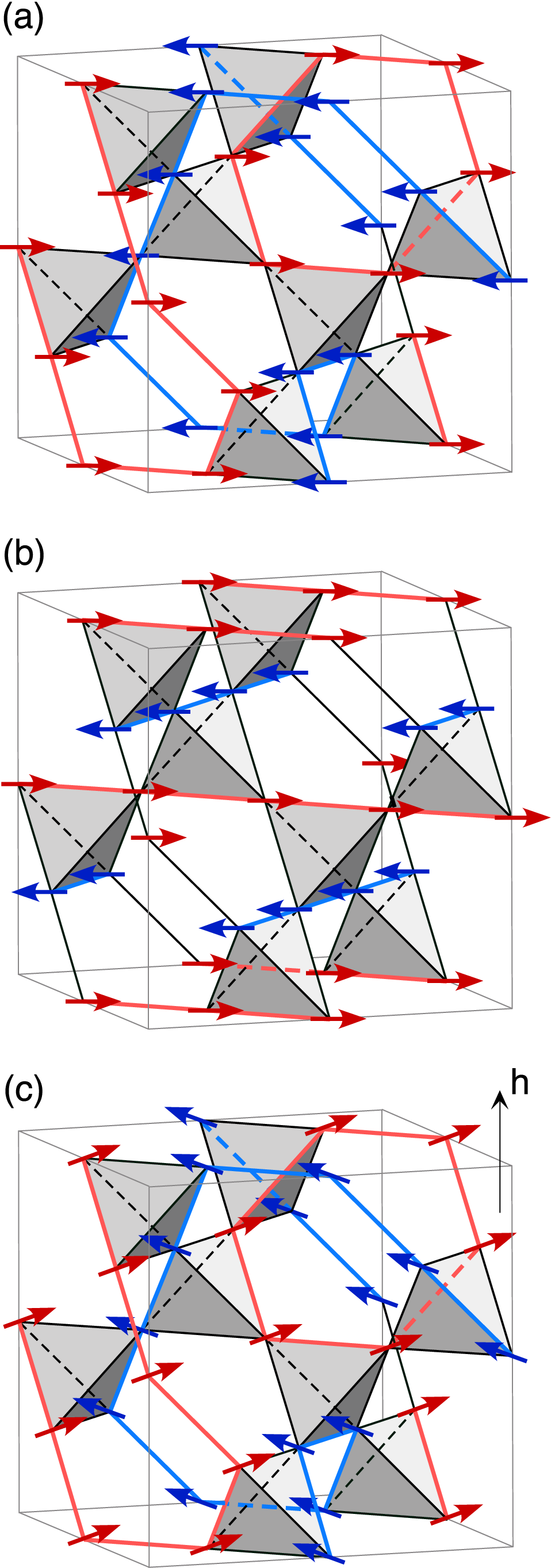}
  \caption{(Color online) (a) Illustration of 
  collinear ${\sf E}$--symmetry nematic state at $h=0$, 
  showing loop--like 
  coordination of parallel spins associated with the ``ice'' manifold.
  (b) Four--sublattice long range order with  ${\sf E}$ symmetry induced by FM $J_3$.
  (c) Canted nematic state with partial magnetization
   under applied field.}
  \label{fig:Enematic}
\end{figure}

\subsection{Collinear nematic phase for $h=0$}

In the absence of magnetic field, the ground state of Eq.~(\ref{eqn:H0})  is determined by the conditions that (i)
the total magnetization of each tetrahedron be zero, to minimize the antiferromagnetic exchange interaction $J_1$,
and (ii) all spins be collinear, to minimize the biquadratic interaction $b$. These conditions select an extensive manifold of  
\begin{eqnarray}
\Omega_0 \approx 1.5^{N/2} \approx 1.22^N
\end{eqnarray}
states with exactly two--``up'' and two--``down'' spins ($uudd$) in each tetrahedron.  
The degeneracy of this ground state manifold  is of the same form as that encountered in Pauling's theory of water ice~\cite{pauling},
and we therefore refer to it as the ``ice'' manifold below.   Since each spin is shared by two neighboring tetrahedra, ``up'' and ``down'' 
spins form unbroken loops as shown in Fig.~\ref{fig:Enematic}.   We return to this point below.

The fact that the direction along which ``up'' and ``down'' spins point is not determined by the Hamiltonian implies that spin 
rotational symmetry must be broken spontaneously (for simplicity, we none-the-less to use ``up'' and ```down'' to denote 
the oppositely oriented spins).   This can be seen in the spin collinearity Eq.~(\ref{eq:collineaity}), which takes on the 
maximal value $P({\bf r})= 1$ for all states in the ice manifold, 
implying that the ground state manifold has nematic (i.e., quadrupolar) order.  
(This is explicitly confirmed by MC simulations below.)
However, as already stated, the ground state manifold does not possess N\'eel order of {\it any} form.

In fact it is possible to calculate the asymptotic form of spin--spin correlations  in the ice manifold by mapping them 
onto configurations of a notional electric (or magnetic) field~\cite{hermele04,izakov04,henley05}.
The condition that every tetrahedron has exactly two--``up'' and exactly two--``down'' spins translates into 
a zero divergence condition for the electric (magnetic) field, and spin--spin correlations
take on a dipolar form
\begin{eqnarray}
\langle {\bf S}_i \cdot {\bf S}_j \rangle \sim \frac{1}{|{\bf r}_i - {\bf r}_j|^3} \, ,
\label{eq:dipolar}
\end{eqnarray}
 dictated by this effective electrodynamics.    
This power--law decay of spin correlations is a signal property of  the ``ice'' manifold.  
However it should not be taken to imply that ``up'' and ``down'' spins are entirely uncorrelated.  
Within each $uudd$ tetrahedron, each spin has twice as many 
AF aligned neighbors as FM aligned ones, and the net correlation on nearest neighbor bonds is 
\begin{eqnarray}
\langle {\bf S}_i. \cdot {\bf S}_j \rangle_{\rm n.n.} = -\frac{1}{3} \, .  
\end{eqnarray}
Locally, order is well formed.   
More formally, we can state that these $uudd$ tetrahedra belong to the two--dimensional ${\sf E}$ irrep of the 
tetrahedral symmetry group ${\cal T}_d$, defined in Sec.~\ref{orderparameters}, 
and that local fluctuations of order $\lambda^{\sf local}_{{\sf E}}$ take on their maximal value 
$\lambda^{\sf local}_{{\sf E}} = 16/3$.

\begin{figure}[t]
  \centering
  \includegraphics[width=7truecm]{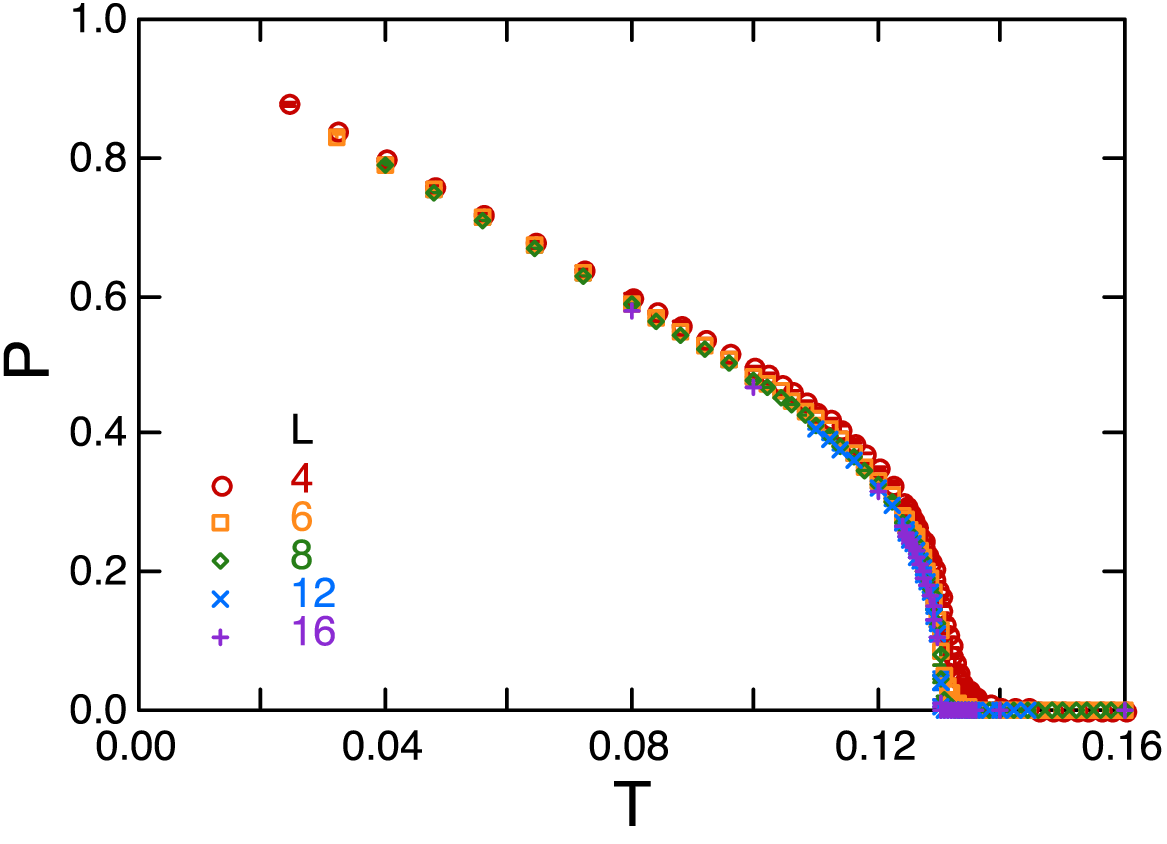}
  \caption{(Color online) 
  Spin collinearity $P$ [Eq.~(\ref{eq:collineaity})] 
  at $h=0$ for \protect\mbox{$b=0.1$}, showing 
  the onset of nematic order at $T_Q \approx 0.13$. 
  $P$ is measured for the farthest spin pair along the $\langle 110 \rangle$ chains 
  in the pyrochlore lattice in each system size ranging from  $L=4$ to $L=16$, 
  and averaged over the $\langle 110 \rangle$ chains 
  running in different directions. 
  \label{fig:spin_collinearity_h=0}}
\end{figure}

\begin{figure}[t]
  \centering
  \includegraphics[width=8truecm]{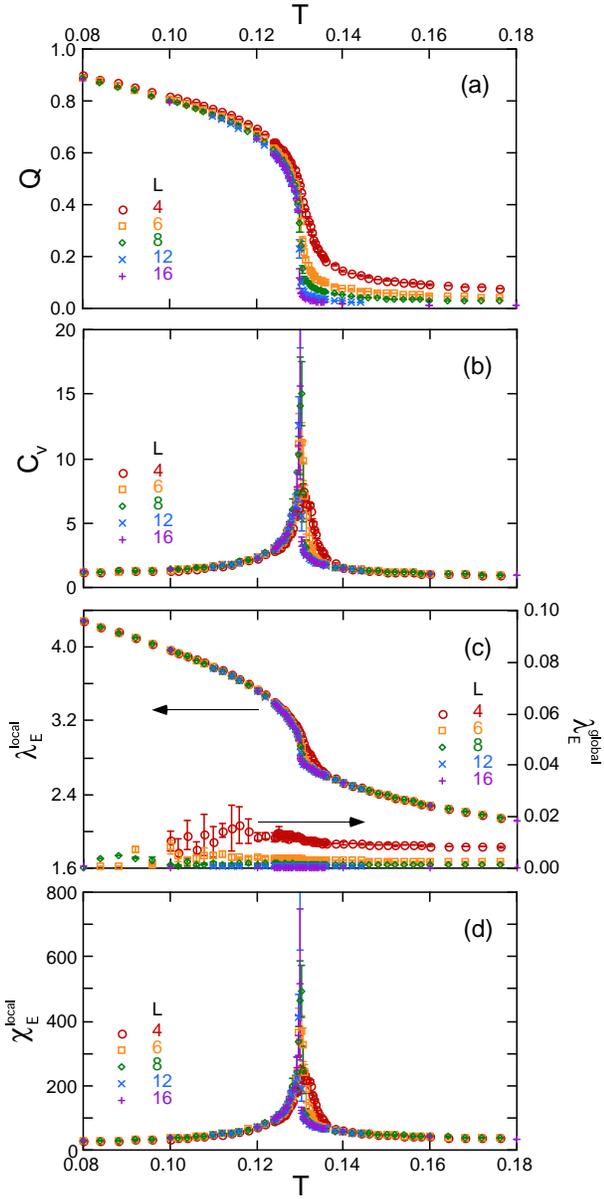}
  \caption{(Color online) 
 (a) Nematic order parameter $Q$ [Eq.~(\ref{eq:Q})], showing 
  the onset of nematic order at $T_Q \approx 0.13$; 
  (b)  Heat capacity;  
(c) Absence of long range four--sublattice order $\lambda^{\sf global}_{{\sf E}}$ is accompanied by well--formed local correlations $\lambda^{\sf local}_{{\sf E}}$;  
(d)  The associated local susceptibility shows a sharp jump at $T_Q$, where spins in tetrahedra with preformed local order gain energy by selecting long range collinearity.
All data are for $h=0$, $b=0.1$, and for system size  ranging from  $L=4$ to $L=16$.  
  \label{fig:b=0.1_J3=0.0_h=0.0}}
\end{figure}

\begin{figure}[t]
  \centering
  \includegraphics[width=7truecm]{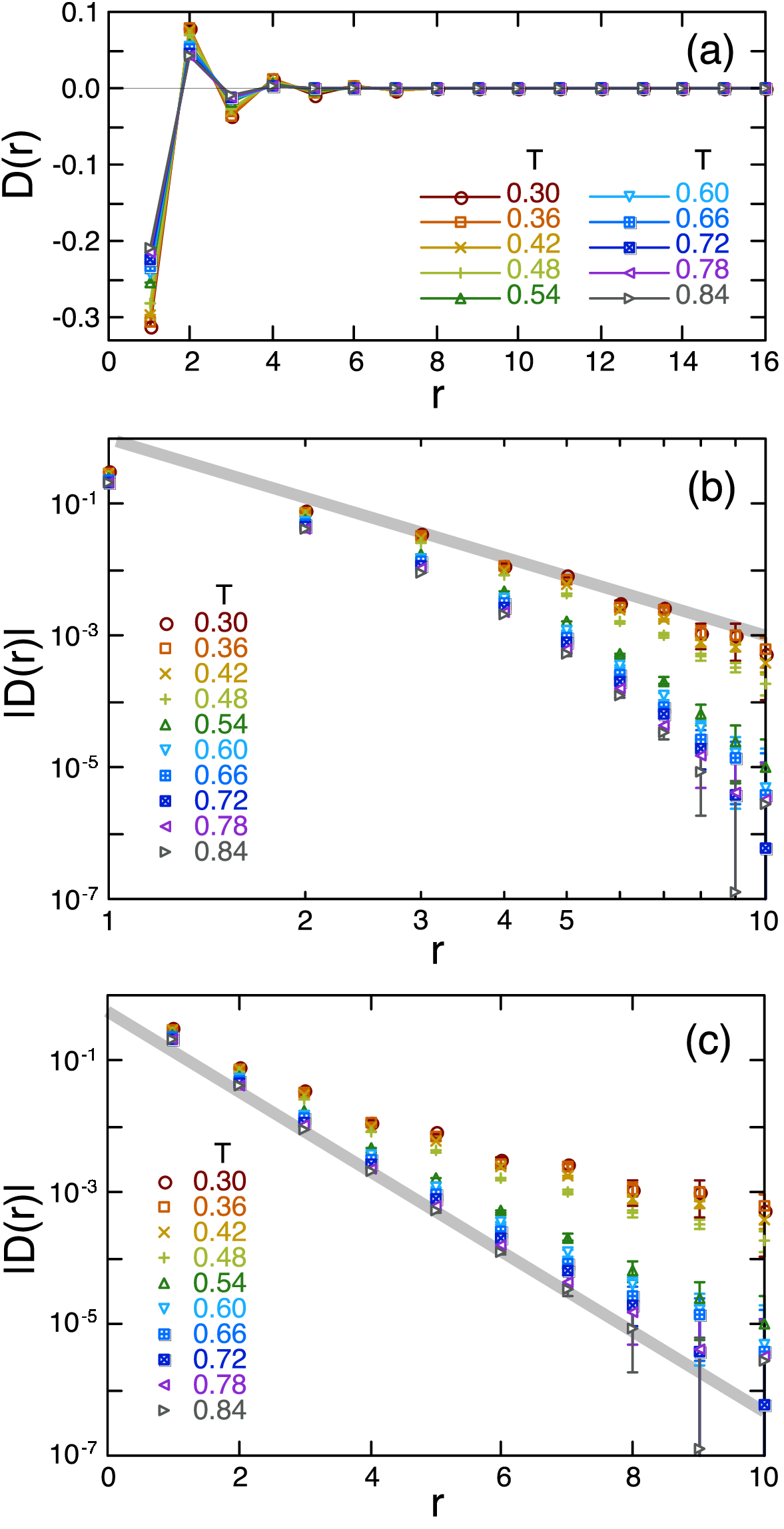}
  \caption{(Color online)  
  (a) Spin correlations $D$ [Eq.~(\ref{eq:spin-spin})]
  at $h=0$ for $b=0.6$, for $r \le L$
  measured along $\langle 110 \rangle$ chains in the pyrochlore lattice and
  averaged over different chain directions as in Fig.~\ref{fig:spin_collinearity_h=0}.
  $r$ is in units of the distance between nearest neighbor spins.
  \label{fig:spin_correlation_h=0}
  (b) Characteristic $1/r^3$ power--law decay of spin correlations 
  in the low--temperature nematic phase for $T < T_Q \sim 0.5$ 
  associated with  the ice manifold of 2:2 states,  
  plotted on a log--log scale. 
  \label{fig:power_law_h=0}
  (c) Exponential decay of spin correlations 
  in the high--temperature paramagnetic phase for $T > T_Q \sim 0.5$,  
  plotted on a log--linear scale. 
  \label{fig:exponential_decay_h=0}
  In (b) and (c), grey lines are guides for the eye, 
  $\sim 1/r^3$ and $\sim \exp(-r)$, respectively.
  In all figures the data are for temperatures ranging from $T=0.30$ to $T=0.84$
  and the system size $L=16$.
  }
\end{figure}

\begin{figure}[t]
  \centering
  \includegraphics[width=7truecm]{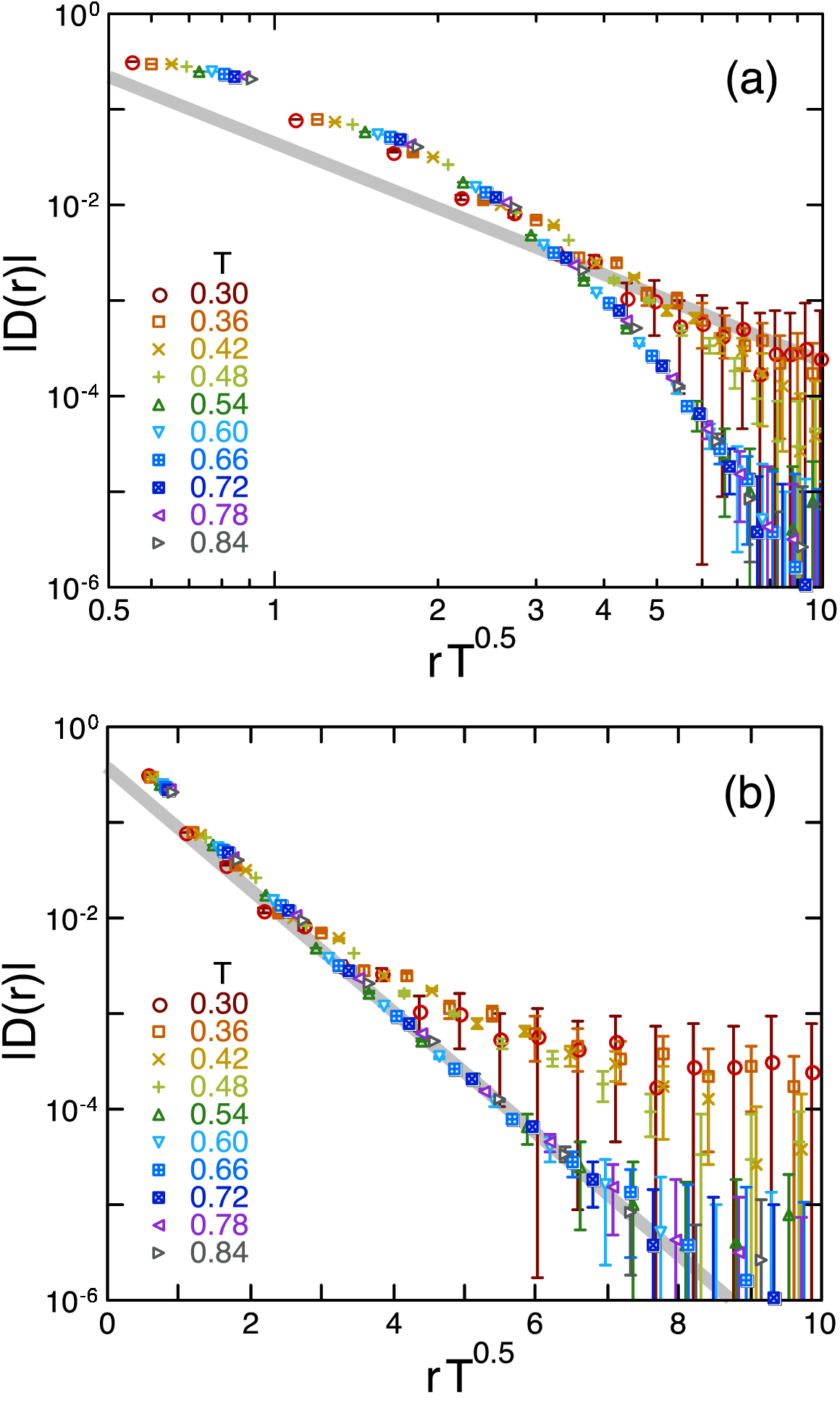}
  \caption{(Color online)  
  Spin correlations spanning the nematic state and high temperature paramagnet plotted as a function of the 
  rescaled distance $r \sqrt{T}$ on (a) log--log scale and (b) log--linear scales.  
  The data are identical to those plotted in Figs.~\ref{fig:power_law_h=0}(b) and (c);  
  grey lines are guides for the eye, showing $\sim 1/r^3$ and $\sim \exp(-r)$, respectively.
  Temperatures should be compared with the nematic ordering temperature $T_Q \approx 0.5$.
   \label{fig:Dr_rescaled_h=0.0}
  }
\end{figure}

This concludes our survey of symmetry breaking for $T=0$, but it leaves open the question, 
what happens at finite temperature?  By analogy with ordered systems where 
``order from disorder'' is effective, thermal fluctuations might be expected to select a 
single configuration from the ``ice'' manifold, and so restore N\'eel order.  
To address this question, we have performed extensive Monte Carlo simulations
using a local--update Metropolis algorithm 
to sample spin configurations.
We typically perform $10^6$ MC samplings for measurements
after $10^5$ steps for thermalization.
We have checked the convergence of the results
by comparing those for different initial spin configurations. 
In particular, to minimize the hysteresis associated with first order transitions, 
we used mixed initial conditions in which different parts of the system are assigned
different ordered or disordered states~\cite{Ozeki2003}. 
Where the acceptance rate in MC updates becomes extremely slow, 
we used the exchange MC method~\cite{Hukushima1996}, 
to avoid local spin-freezing at low temperatures. 
Results are divided into five bins to estimate statistical errors
by variance of average values in the bins.
The system sizes in the present work are up to $L=16$,
where $L$ is the linear dimension of the system
measured in the cubic units shown in Fig.~\ref{fig:pyrochlore}, i.e.,
the total number of spins $N$ is given by $16L^3 $.  
We show the results for $b=0.1$ and $b=0.6$, case by case,
both of which exhibit qualitatively the same behavior.

In the absence of applied magnetic field, the spin collinearity $P({\bf r})$ 
grows sharply below a transition temperature \protect\mbox{$T_Q \approx b$},
as illustrated in Fig.~\ref{fig:spin_collinearity_h=0}.  
As expected, for $T \to 0$, $P({\bf r}) \to 1$, implying that all spins have a single common axis.   
The nematic order parameter ${\it Q}$ [Eq.~(\ref{eq:Q})]
is plotted in Fig.~\ref{fig:b=0.1_J3=0.0_h=0.0}(a), 
together with the heat capacity in Fig.~\ref{fig:b=0.1_J3=0.0_h=0.0}(b) for a range 
of system sizes from $L=4$ to $L=16$.   
Here the heat capacity is calculated by the fluctuation of internal energy as
\begin{equation}
C_v = \frac{\langle {\cal H}^2 \rangle - \langle {\cal H} \rangle^2}{T^2 N} \, .
\label{eq:Cv}
\end{equation}
The sharp onset of order 
and jump in heat capacity 
imply a first order phase transition at $T_Q \approx 0.13$.

Treating this nematic order at the level of a Ginzburg-Landau theory, the free-energy terms allowed by lattice and 
spin rotational symmetries are
\begin{eqnarray}
\label{eqn:nematicF}
 \mathcal{F} & =&  a Q^2 + b Q^3 + c Q^4 + \ldots \, ,
\end{eqnarray}
where $Q^2$ is defined in Eq.~(\ref{eq:Q}), the third order invariant $Q^3$ is given by 
\begin{eqnarray}
Q^3 &= & 2 ({Q^{3 z^2-r^2}})^3 + 3 Q^{3 z^2-r^2} \left[(Q^{x z})^2 + (Q^{y z})^2 \right] \nonumber\\
&&- 6 Q^{3 z^2-r^2} \left[(Q^{x^2-y^2})^2 + (Q^{x y})^2 \right]   \nonumber\\
&&+ 3 \sqrt{3} \Big( Q^{x^2-y^2} \left[(Q^{x z})^2 -  (Q^{y z})^2\right] 
\nonumber\\
&& \qquad \qquad + 
2 Q^{x y} Q^{x z} Q^{y z} \Big) \, .
\end{eqnarray}
In a three--dimensional uniaxial nematic state, such as that realized here, all quadrupoles moments 
$Q^n$ are proportional to a simple scalar $Q$.  The presence of a cubic term in the free energy Eq.~(\ref{eqn:nematicF}) therefore implies that the phase transition from nematic phase to paramagnet as a function of temperature must be first order --- as observed in the MC results.

In principle, thermal fluctuations might select a single N\'eel state from the ice manifold, in which case $T_Q$ would mark 
the onset of dipolar as well as quadrupolar order.   However this is not the case.  
Spin--spin correlations $D({\bf r})$, defined in Eq.~(\ref{eq:spin-spin}), remain short ranged [Fig.~\ref{fig:spin_correlation_h=0}(a)].
At the distances accessible to simulation, they rapidly cross over from the power--law decay characteristic of the ice manifold
at low temperatures [Fig.~\ref{fig:power_law_h=0}(b)] 
to the exponential decay expected for a paramagnet 
[Fig.~\ref{fig:exponential_decay_h=0}(c)].

The reason that the usual order--from--disorder mechanism is ineffective in selecting dipolar order is the massive degeneracy of
the ice manifold --- the entropy gain of fluctuations about the favored state (relative to the average) would have 
to compensate for the loss of an extensive entropy of 
\begin{eqnarray}
\ln \Omega_0 / N \sim 0.5 \ln 1.5 \approx 0.20 \nonumber
\end{eqnarray}
per spin.  We return to this point below.

At finite temperature, the algebraic decay of spin correlations $D(r) \sim 1/r^3$  in the Coulomb phase is expected to crossover to 
exponential decay $D(r) \sim \exp (-r/\xi_c)$ for $r \gtrsim \xi_c$, where the characteristic length scale $\xi_c$ diverges for $T \to 0$.    
For Heisenberg spins in three dimensions, $\xi_c \sim 1/\sqrt{T}$~\cite{henley05,garnin99,canals01}.  
This is the only length scale in the simplest Coulomb theory, and it is therefore interesting to plot the spin correlations $|D(r)|$ 
for a rescaled distance $r \sqrt{T}$.   This is done in Fig.~\ref{fig:Dr_rescaled_h=0.0}.  
At low temperatures $T < T_Q$, the data appear to collapse onto a single power--law behavior 
in this range, while they collapse onto an exponential behavior above $T_Q$: 
There is a rapid change between these behaviors, associated with the discontinuous transition at $T=T_Q$. 
The results suggest that the Coulomb--phase theory applies to 
the present bilinear--biquadratic model, and in addition, 
that the characteristic length $\xi_c$ suddenly changes from several lattice spacings  
in the nematic phase for $T < T_Q$
to one comparable to the lattice spacing 
in the paramagnetic phase for $T > T_Q$.

Once again, these results have a simple interpretation in terms of local, preformed order.  
In Fig.~\ref{fig:b=0.1_J3=0.0_h=0.0}(c) we plot the expectation value of the order parameter for the simplest kind of four--sublattice order, $\lambda^{\sf global}_{{\sf E}}$ [Eq.~(\ref{eq:global})].   
This clearly scales to zero with system size.   However there is a sharp
feature in the susceptibility associated with $\lambda^{\sf local}_{{\sf E}}$ [Eq.~(\ref{eq:local})] at $T_Q$, shown in 
Fig.~\ref{fig:b=0.1_J3=0.0_h=0.0}(d), where tetrahedra with {\it local} ${\sf E}$ symmetry collectively choose 
collinear configurations.   Indeed, as $T \to 0$, $\lambda^{\sf local}_{{\sf E}}$ takes on its maximum allowed value of 
$\lambda^{\sf local}_{{\sf E}} \to 16/3$ [Fig.~\ref{fig:b=0.1_J3=0.0_h=0.0}(c)], 
as required for loops of perfectly collinear spins.

\subsection{Nematic phase with local ${\sf E}$ symmetry}

\begin{figure}[t]
  \centering
  \includegraphics[width=7truecm]{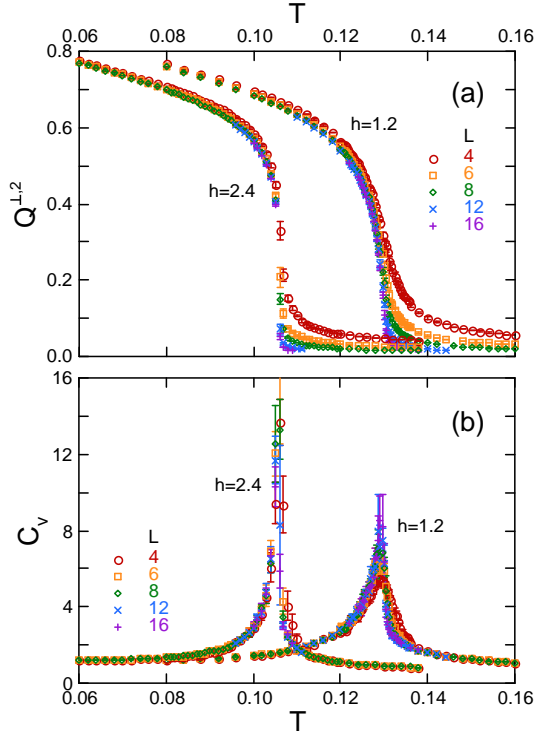}
  \caption{(Color online)  
  (a) Nematic order parameter $Q^{\perp,2} = |{\bf Q}^{\perp,2}|$ [Eq.~(\ref{eq:Qperp2})], 
  signaling the nematic long--range order in finite magnetic field;
  (b) Heat capacity [Eq.~(\ref{eq:Cv})].
  Data are at $h=1.2$ and $h=2.4$ for $b=0.1$, and for system size ranging from $L=4$ to $L=16$.
  \label{fig:b=0.1_J3=0.0_h=1.2_2.4}
  }
\end{figure}

In applied magnetic field, the ``up'' and ``down'' spins of the collinear nematic phase immediately ``flop'' into the plane 
parallel to ${\bf h}$, and transform into the 2:2 canted coplanar configurations shown in Fig.~\ref{fig:tetrahedron}.   Such canting is entirely compatible with the ice manifold, as is illustrated in Fig.~\ref{fig:Enematic}(c) --- entire loops of spins cant simultaneously, to give a state with smoothly evolving magnetization, but {\it no} N\'eel order.   

The correlation function $P({\bf r})$ retains a finite (reduced) value in this new canted manifold of states.  However
spin rotational symmetry is now explicitly broken by the magnetic field, so this does not of itself imply nematic order.   
Nematic order is none the less present, in the selection of a common plane within which the spins cant.   This is equivalent
to the selection of a direction (but not an orientation) in the $xy$ plane, and long range order can now be observed
in the transverse moment ${\bf Q}^{\perp,2}$ defined by Eq.~(\ref{eq:Qperp2}), 
as discussed in Table~\ref{tab:quadrupoles}.

Since this director breaks the residual $O(2)$ symmetry, the resulting nematic state must possess a branch of gapless (Goldstone) 
modes associated with rotations of the plane of canting about the $z$ axis.  
It is worth noting that a canted N\'eel state with ${\sf E}$--type symmetry 
would break rotational symmetry in the same 
way~\cite{motome06}.  However for ${\mathcal H}^{\sf LRO}\to 0$, simulations show that spin--spin correlations 
retain their power--law character at low temperatures, implying the absence of long--range N\'eel order.

The transition from local--${\sf E}$ symmetry nematic state to collinear state as $h \to 0$ is completely smooth, 
and the finite $T$ properties of nematic state at finite $h$ are qualitatively identical to those shown in 
Figs.~\ref{fig:b=0.1_J3=0.0_h=0.0} and \ref{fig:exponential_decay_h=0}, 
with the obvious caveats that $P({\bf r}) < 1 $ for $T \to 0$, 
and the collinear order parameter ${\it Q}$ must be replaced by $Q^{\perp,2} = |{\bf Q}^{\perp,2}|$.  
Once again the onset of nematic order at $T_Q$ is associated with a sharp peak in heat capacity, 
a rise in the local fluctuations with ${\sf E}$ symmetry, $\lambda_{\sf E}^{\sf local}$, 
and the absence of long range order of the form $\lambda_{\sf E}^{\sf global}$.  

A suitable free energy to describe this nematic state is
\begin{eqnarray}
\mathcal{F} & =&  a_2 |{\bf Q}^{\perp,2}|^2  + c_{22}  |{\bf Q}^{\perp,2}|^4 + e_{222} |{\bf Q}^{\perp,2}|^6 \, ,
\end{eqnarray}
which permits both first and second order phase transitions into a paramagnetic phase as a function of temperature, 
depending on the sign of $c_{22}$.  However MC simulations suggest that the transition remains first order.
Figure~\ref{fig:b=0.1_J3=0.0_h=1.2_2.4} shows the temperature dependences of 
the nematic order parameter $Q^{\perp,2} = |{\bf Q}^{\perp,2}|$ [Eq.~(\ref{eq:Qperp2})] and
the heat capacity [Eq.~(\ref{eq:Cv})] for $h=1.2$ and $h=2.4$. 
For both cases, the order parameter exhibits a sharp onset 
and the heat capacity shows a jump, 
indicating that the ${\sf E}$--symmetry nematic transition is of the first order, 
as for $h=0$ in Fig.~\ref{fig:b=0.1_J3=0.0_h=0.0}. 
As noted by comparing the results for $h=1.2$ and $h=2.4$, 
the discontinuity becomes clearer as $h$ increases. 

\begin{figure}[t]
  \centering
  \includegraphics[width=7truecm]{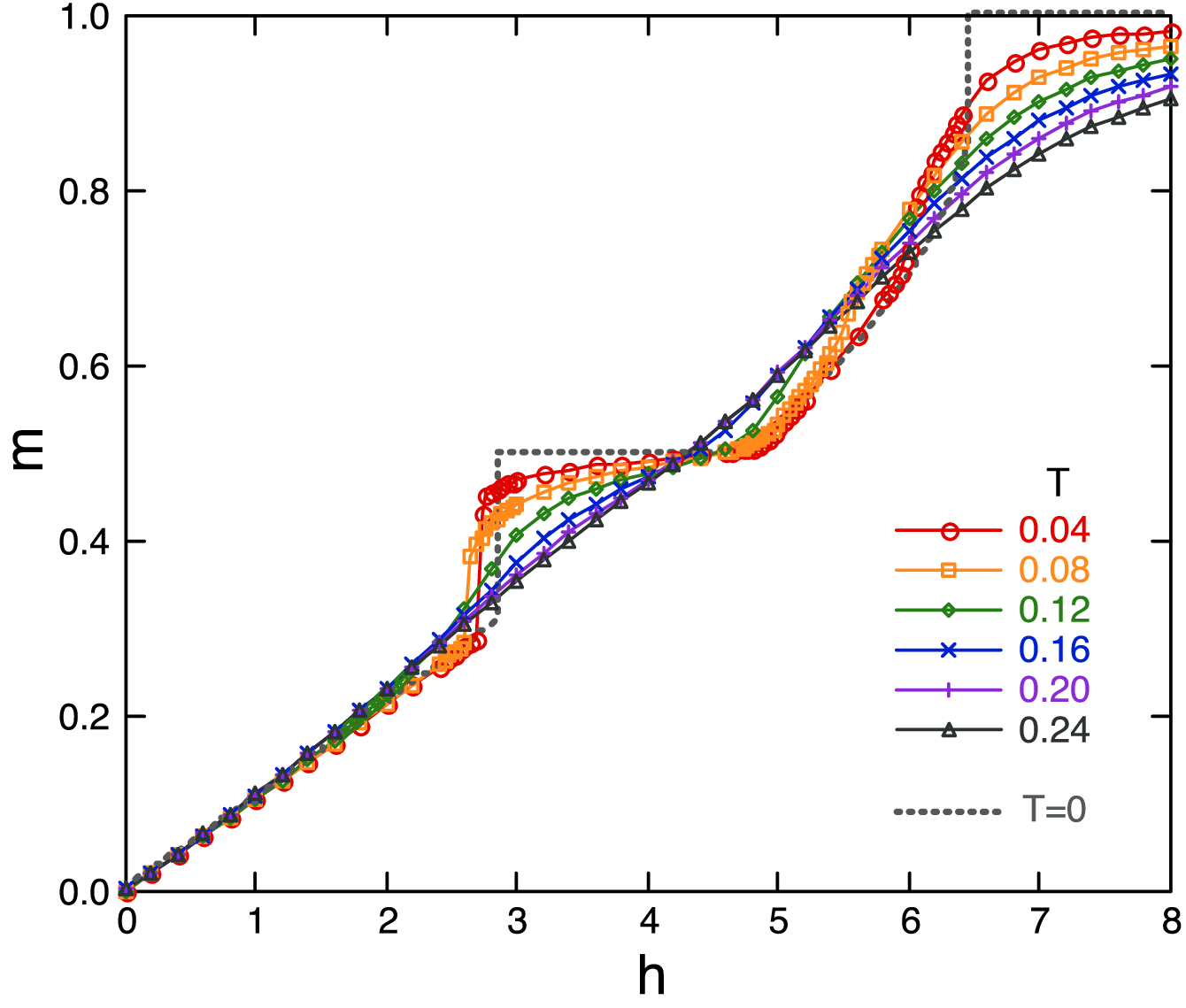}
  \caption{(Color online) Dependence of the magnetization $m$ [Eq.~(\ref{eq:m})]
  on magnetic field $h$ for $b=0.1$ and $J_3=0$, showing the 
  existence of the magnetization plateau in the absence of long--range N\'eel order.   Symbols show 
  the result of Monte Carlo simulations for temperatures ranging from $T=0.04$ to $T=0.24$
  and the system size $L=8$.   The dashed line is the result obtained by minimizing the energy for $T=0$.
\label{fig:mofh}}
\end{figure}

In principle the ${\sf E}$--symmetry nematic state could interpolate to saturation, simply by canting all spins until 
they are aligned with the magnetic field.   However this is not energetically favorable at the level of a single 
tetrahedron (Fig.~\ref{fig:tetrahedron}), and for a magnetic 
field $h \approx 3$, the system undergoes a first order transition into a state with magnetization $m=1/2$, 
seen as the plateau in Fig.~\ref{fig:mofh}.  This state is discussed in detail in the section below.

%%%%%%%%%%%%%%%%%%%%%%%%%%%%%%%%%%%%%%%%%%%%

\subsection{Plateau liquid with local ${\sf T_2}$ symmetry}
\label{plateauliquid}

\begin{figure}[t]
  \centering
  \includegraphics[width=5.5truecm]{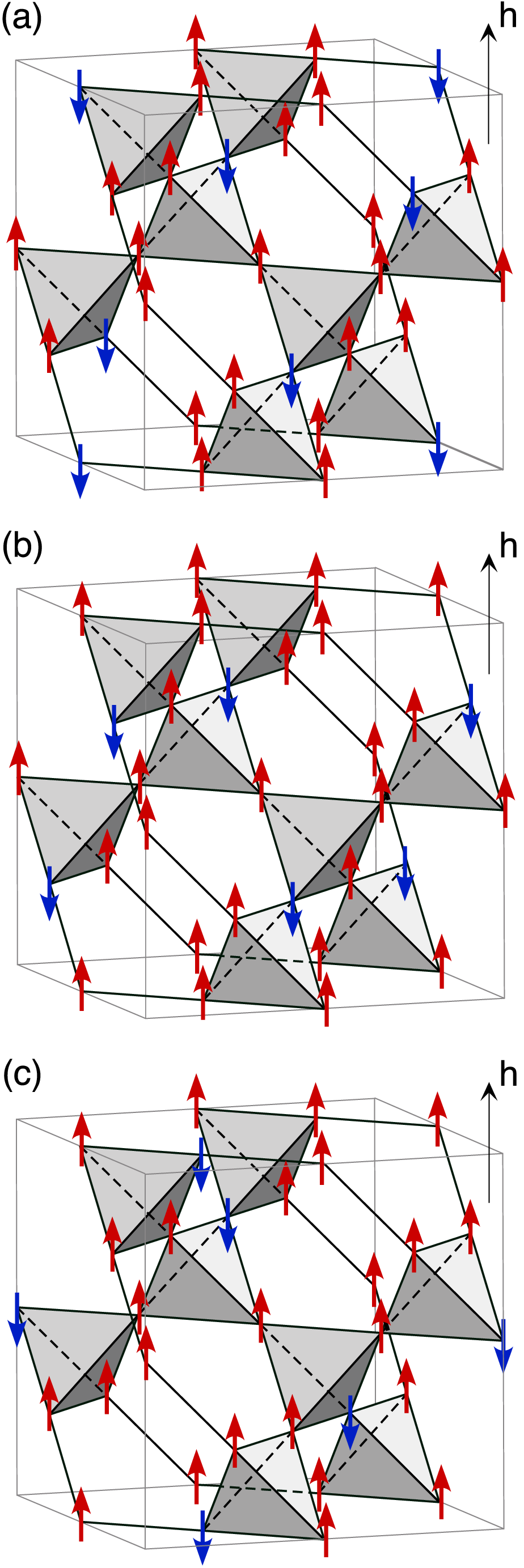}
  \caption{(Color online) 
  Half--magnetization plateau states ($uuud$ states) on a pyrochlore lattice 
  with exactly  three--up and one--down spins per tetrahedron.
  (a) Schematic picture of an $uuud$ state with no long range order, 
  associated with the ``dimer" manifold;
  (b) $uuud$ state with long--range four--sublattice order with ${\sf T_2}$ symmetry 
  induced by FM $J_3$, as considered in 
  Refs.~[\protect\onlinecite{penc04}] and [\protect\onlinecite{motome06}]; 
  (c) 16--sublattice order. See also Fig.~\ref{fig:lnDetM_Nflip}.
  \label{fig:liquid-solid2}}
\end{figure}

The half--magnetization plateaux observed in Cr spinels are associated with collinear states with three--up 
and one--down spin per tetrahedron.   There are in fact an extensive number 
\begin{eqnarray}
\Omega_0 \approx 1.7^{N/4} \approx 1.14^N
\end{eqnarray}
of such $uuud$ states ---  a manifold isomorphic to hard--core dimer coverings of the 
diamond lattice formed by joining the centers of tetrahedra\cite{Nagle}.    
(Dimers on bonds of the diamond lattice correspond to the down spins in $uuud$ states on the pyrochlore lattice.)
We therefore refer to it as the ``dimer'' manifold below.  Collinear $uuud$ states with and without simple N\'eel order
are illustrated in Fig.~\ref{fig:liquid-solid2}.

\begin{figure}[t]
  \centering
  \includegraphics[width=7truecm]{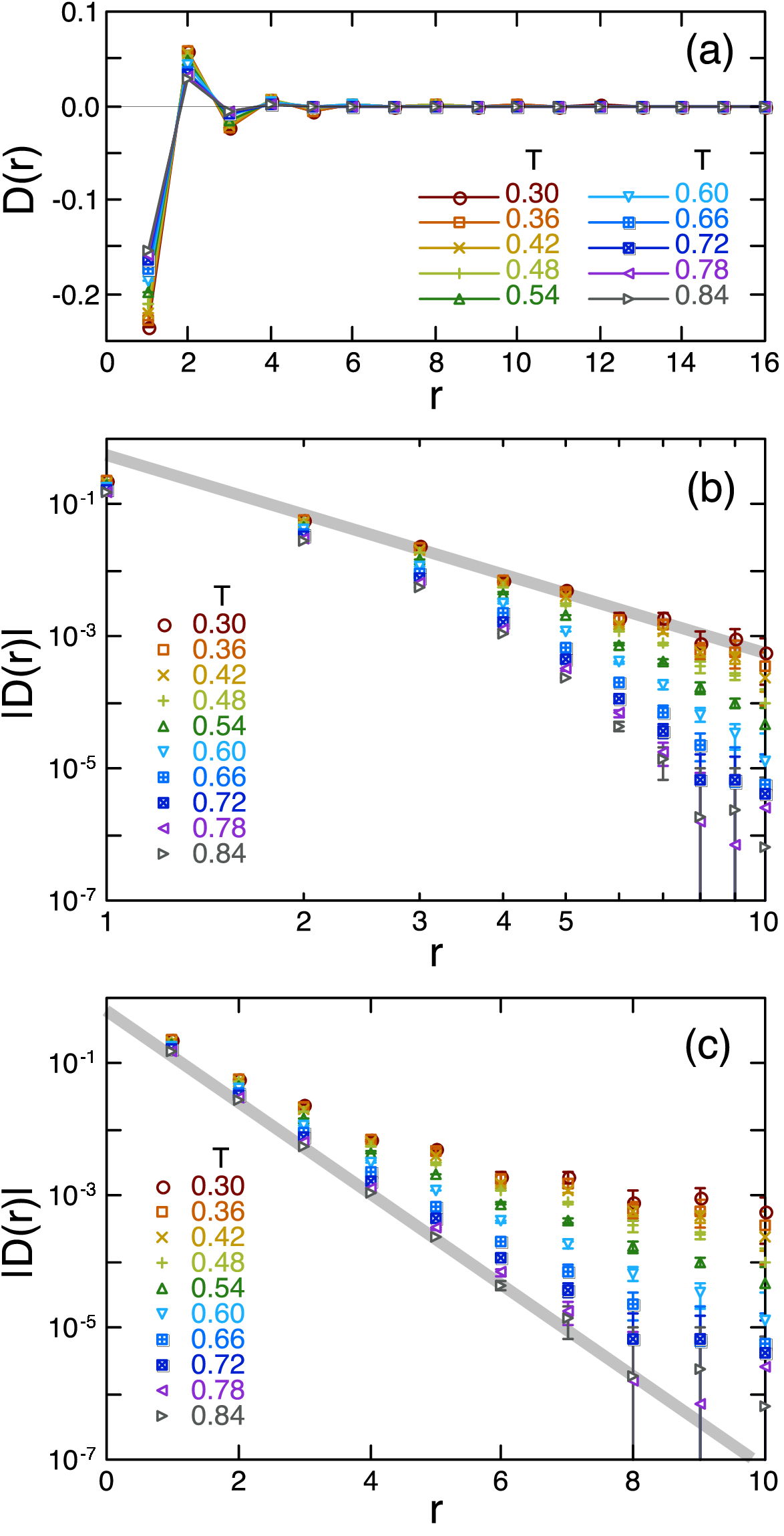}
  \caption{(Color online) Absence of long--range magnetic order in the
  plateau liquid state for $b=0.6$, $J_3=0$, and $h=4$.
  (a) The reduced spin correlation function $D(r)$, defined 
  by~\protect\mbox{Eq.~(\ref{eq:spin-spin})}, 
  measured along the $\langle 110 \rangle$ chains, as in Fig.~\ref{fig:spin_correlation_h=0}(a). 
  (b) Characteristic $1/r^3$ power--law decay of spin correlations at low temperatures 
  $T < T^* \sim 0.6$, 
  associated with the dimer manifold of $uuud$ states, 
  plotted on a log--log scale. 
  (c) Exponential decay of spin correlations at high temperatures $T > T^* \sim 0.6$, 
  plotted on a log--linear scale. 
  In (b) and (c), grey lines show guides for the eye, 
  $\sim 1/r^3$ and $\sim \exp(-r)$, respectively.
  In all figures the data are for temperatures ranging from $T=0.30$ to $T=0.84$ 
  and the system size $L=16$. 
 \label{fig:liquid}}
\end{figure}

\begin{figure}[t]
  \centering
  \includegraphics[width=7truecm]{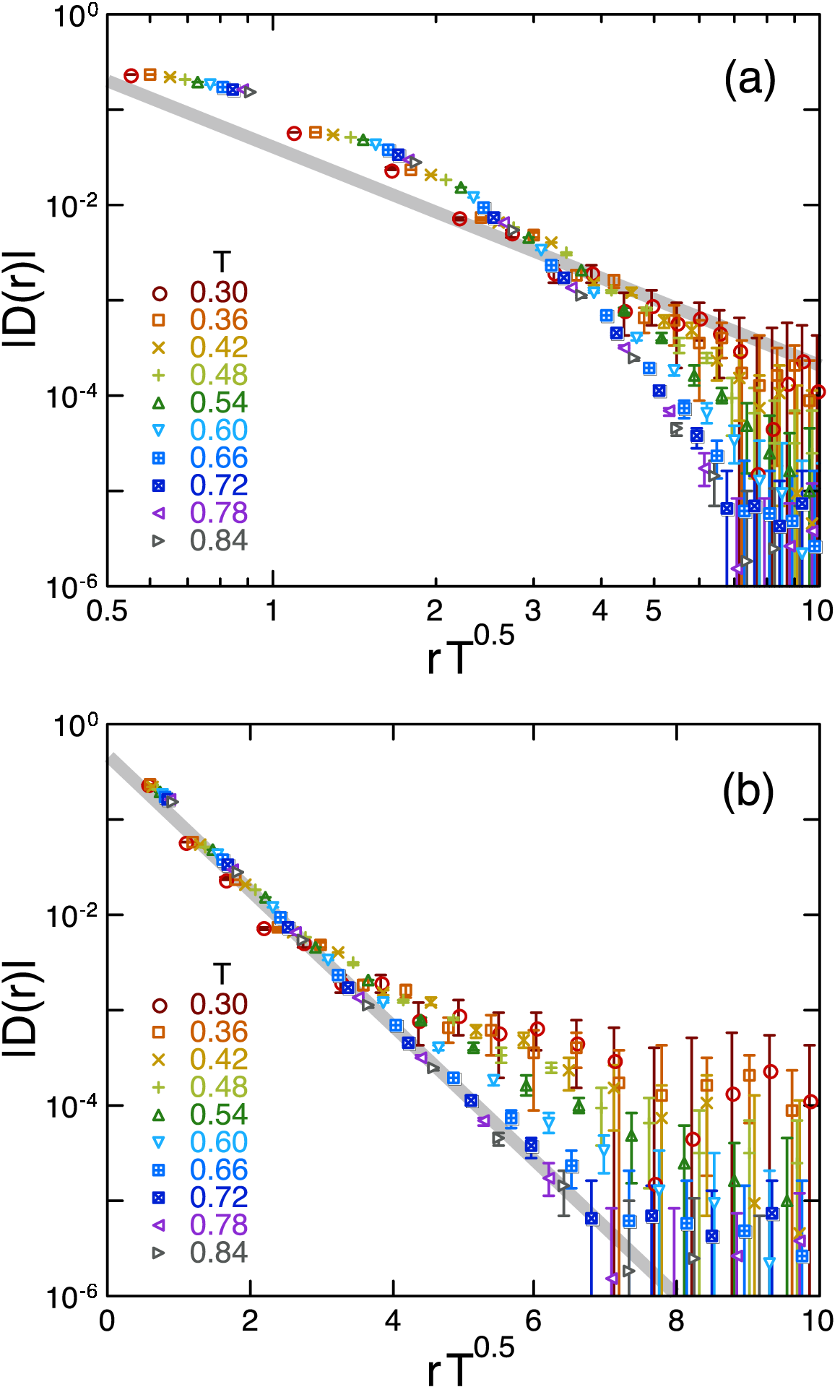}
  \caption{(Color online) 
  Spin correlations spanning the plateau liquid and high temperature paramagnetic phases, 
  plotted as a function of the rescaled distance $r \sqrt{T}$ on (a) log--log scale and 
  (b) log--linear scales.  
  The data are identical to those plotted in Figs.~\ref{fig:liquid}(b) and (c); 
  grey lines are guides for the eye, showing $\sim 1/r^3$ and $\sim \exp(-r)$, respectively.
  Temperatures should be compared with the crossover scale $T^* \approx 0.6$.  
  \label{fig:Dr_rescaled_h=4.0}
  }
\end{figure}

\begin{figure}[t]
  \centering
  \includegraphics[width=8truecm]{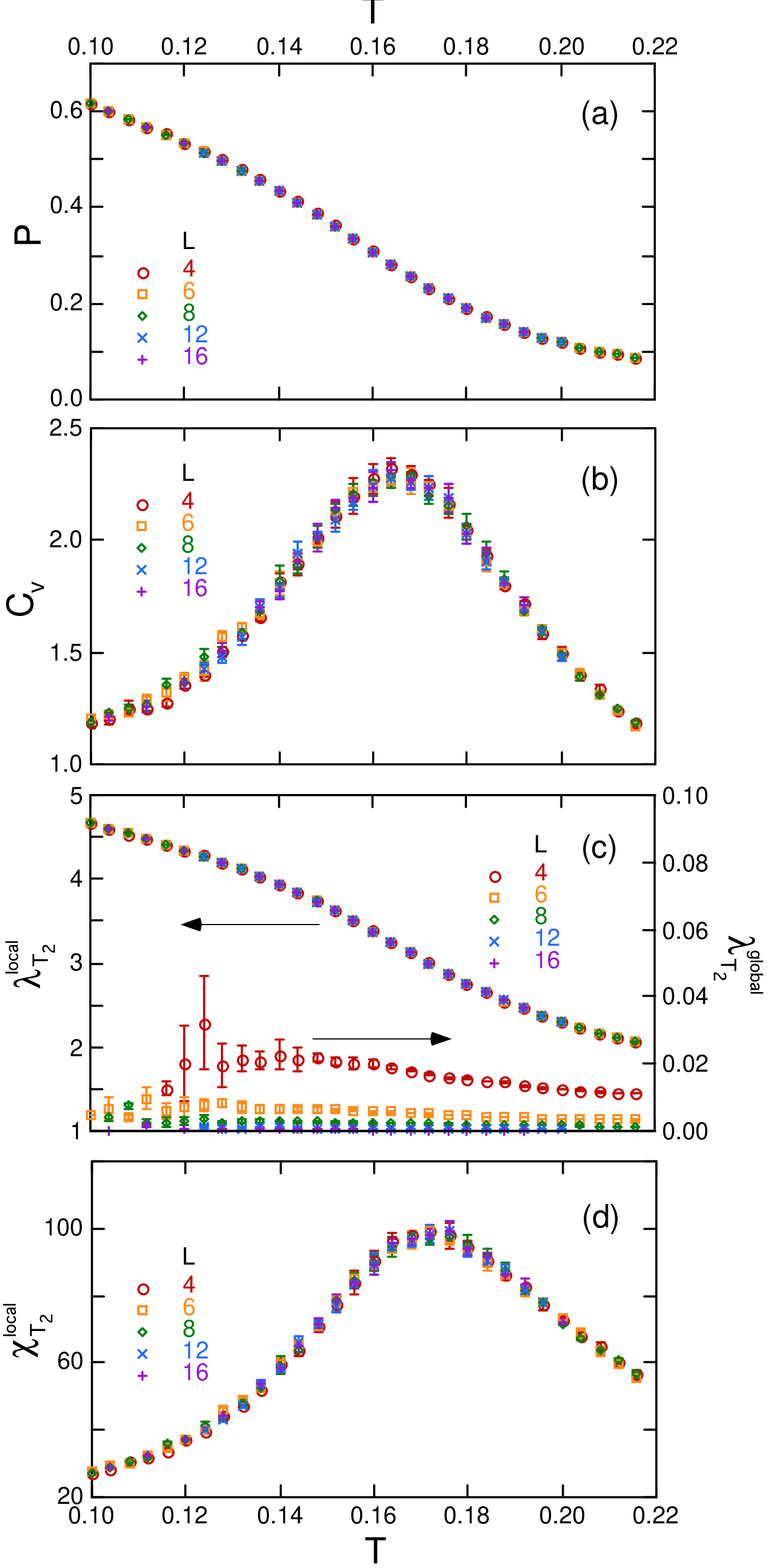}
  \caption{(Color online) Temperature dependence of 
  (a) collinearity [\protect\mbox{Eq.~(\ref{eq:collineaity})}, 
  cf. \protect\mbox{Fig.~\ref{fig:spin_collinearity_h=0}}], 
  (b) heat capacity [\protect\mbox{Eq.~(\ref{eq:Cv})}], 
  (c) the related measure of {\it local} 
  correlation $\lambda_{{\sf T_2}}^{\sf local}$ defined by \protect\mbox{Eq.~(\ref{eq:local})}, 
  and the  {\it global} order parameter $\lambda_{{\sf T_2}}^{\sf global}$ 
  defined by \protect\mbox{Eq.~(\ref{eq:global})}, and
  (d) the associated local susceptibility.  Simulations were performed for 
  $h=4$, $b=0.6$, in clusters with $L=4$ to $L=16$. 
  \label{fig:chi}}
\end{figure}

It is possible to construct a field theory for the dimer manifold at $T=0$ by exact analogy 
with the treatment of the ice manifold above.   The condition that every tetrahedron has exactly three--up 
and exactly one--down spins translates into a zero divergence condition for an electric (magnetic) field, 
modified to include a source term~\cite{bergman06}.

Once again, thermal fluctuations are ineffective in restoring long--range N\'eel order.
Reduced spin--spin correlations Eq.~(\ref{eq:spin-spin}) at finite distance exhibit a crossover between a dipolar form 
[cf. Eq.~(\ref{eq:dipolar})] for low temperatures, and exponential decay for high temperatures 
--- see Fig.~\ref{fig:liquid}.   While spins are perfectly collinear at low temperatures [Fig.~\ref{fig:chi}(a)], 
the $z$ axis is now singled out by magnetic field, and $Q^{3z^2-r^2}$ contributes to $P({\bf r})$.

This means that there is no symmetry breaking associated with the smooth rise in collinearity for
$T^* \approx b$, which should be regarded as a crossover rather than a phase transition.
Singular features are similarly absent from the  heat capacity, shown in Fig.~\ref{fig:chi}(b).  
This smooth change is also seen in the rescaled plot of the spin correlations 
shown in Fig.~\ref{fig:Dr_rescaled_h=4.0}.  
The crossover from the low--$T$ power--law behavior to the high--$T$ exponential decay 
is much more smooth compared to the case for the nematic transition at $h=0$ 
in Fig.~\ref{fig:Dr_rescaled_h=0.0}.
This suggests a smooth growth of the characteristic length scale 
$\xi_c \sim 1/\sqrt{T}$ at $T \sim T^* \approx b$. 
We therefore conclude that the magnetization plateau is a true spin--liquid state, 
continuously connected with the high--$T$ paramagnet.   We refer to this as the plateau liquid below.

It is interesting to note that, despite the absence of {\it any} kind of long range order, the defining 
property of the plateau liquid --- its magnetization (Fig.~\ref{fig:mofh}) --- 
is almost indistinguishable from those of the four--sublattice ordered state~\cite{motome06}.
Long--range four--sublattice order $\lambda_{{\sf T_2}}^{\sf global}$ is {\it explicitly} 
absent --- the plateau liquid possess the full ${\sf A_1}$ symmetry of the paramagnet.  
None the less there is a marked rise in local ${\sf T_2}$ order of individual 
tetrahedra $\lambda_{{\sf T_2}}^{\sf local}$ at $T^* \approx b$
accompanied by a broad peak in its susceptibility, 
as shown in Figs.~\ref{fig:chi}(c) and (d). 
This curious spin--liquid state clearly deserves further study.

To this end, we have performed low--$T$ expansions of the free energy of 
many different ordered and disordered $uuud$ states.   These are controlled expansions
about the ground state in powers of $T$ for a spin of length $S=1$, where we write the energy
\begin{equation}
{\mathcal H} = 
E_0 +  \frac{1}{2} \sum_{i, j} \delta S_i  {\cal M}_{ij} \delta S_j
+ \ldots \, ,
\end{equation}
in terms of the fluctuations 
\begin{eqnarray}
{\bf \delta S} = \left(\delta S^x_1,\delta S^x_2, \ldots \delta S^x_N, \delta S^y_1,\delta S^y_2,  \ldots \delta S^y_N \right)
\end{eqnarray}
about a given $uuud$ configuration.    The leading fluctuation contribution to the free energy 
can then be calculated in terms of the trace over eigenvalues of the $2N\times 2N$ matrix ${\cal M}$
in the form
\begin{eqnarray}
\frac{{\mathcal F}}{N}
&=&  \frac{E_0}{N}
 - T \ln T + 
 \frac{T}{2 N} \langle \ln \det{\cal M} \rangle_{\Omega_0}  \nonumber\\
&& \quad - \frac{T}{N} \ln \Omega_0  + {\mathcal O}(T^2),
\end{eqnarray}
where $E_0$ is the ground state energy, $\Omega_0$ its degeneracy, 
and $\langle ... \rangle_{\Omega_0}$ the average over all degenerate ground states.

For a generic ordered phase, $\Omega_0$ is finite, and $\det{\cal M}$ takes on
the same value for all (symmetry related) ground states.    
In this case $\ln \Omega_0/N \to 0$ for $N \to \infty$.  
However for the dimer manifold, $\Omega_0 \approx 1.14^{N}$, which means
that the ground state has a {\it finite entropy per site} 
\begin{equation}
\frac{S_0}{N} \approx \ln 1.14 \approx 0.13 \;.
\end{equation}
In this case, different ground states are not related by simple lattice symmetries 
and the fluctuation entropy per site
\begin{eqnarray}
s_{\sf f} = - \frac{\ln \det{\cal M}}{2N}
\label{eq:s_f}
\end{eqnarray} 
takes on a range of values.   

We have studied the distribution of values of $s_{\sf f}$ within the dimer manifold for
a range of values of $b$, by numerically calculating $\det{\cal M}$ for 
$10000$ randomly generated $uuud$ states in a cluster of $N=1024$ sites ($L=4$), using a Monte Carlo algorithm based on loop updates of spins.  
We found that the highest value of $s_{\sf f}$ is achieved by an eight--fold degenerate, 16--sublattice ``{\sf R}-state'' \cite{bergman06}, in 
which the four A--sublattice tetrahedra within the 16--site cubic unit cell of the pyrochlore lattice take on all four possible 
$uuud$ configurations [Fig.~\ref{fig:liquid-solid2}(c)].   This state has overall cubic symmetry, and is actually observed in the 
plateau phase of HgCd$_2$O$_4$~\cite{matsuda07}.   
The lowest value of $s_{\sf f}$ is achieved by the four--sublattice order shown in Fig.~\ref{fig:liquid-solid2}(b). 
The calculated values of the maximum and minimum values $s_{\sf f}^{\sf max}$ and $s_{\sf f}^{\sf min}$ are 
listed in Table~\ref{tab:s_f} 
together with the mean value $\langle s_{\sf f} \rangle$ and 
the difference between $s_{\sf f}^{\sf max}$ and the mean $\langle s_{\sf f} \rangle$, $\Delta s_{\sf f}$.

\begin{table} 
\caption{\label{tab:s_f} Fluctuation entropy per site calculated for randomly generated $uuud$ states in an $N=1024$ cluster.
Here $s_{\sf f}^{\sf min}$, $s_{\sf f}^{\sf max}$, and $\langle s_{\sf f} \rangle$
are the lowest, highest, and mean value of the entropy, respectively, and 
$\Delta s_{\sf f} = s_{\sf f}^{\sf max} - \langle s_{\sf f} \rangle$ measures the deviation of the highest 
value of entropy from the mean.    Statistical errors on all numbers are less than $10^{-6}$.}
\begin{ruledtabular} 
\begin{tabular}{cccccc} 
$b$   & $s_{\sf f}^{\sf min}$  & $s_{\sf f}^{\sf max}$  & $\langle s_{\sf f} \rangle$   & $\Delta s_{\sf f}$ \\
\hline
\hline
0.05 & -0.79931 & -0.79608 & -0.79837 & 0.00228\\
0.1 & -1.63451 & -1.63191 & -1.63374 & 0.00183\\
0.2 & -2.54944 & -2.54758 & -2.54888 & 0.00130\\
0.3 & -3.12864 & -3.12725 & -3.12823 & 0.00098\\
0.4 & -3.56055 & -3.55948 & -3.56025 & 0.00077\\
0.5 & -3.90757 & -3.90672 & -3.90734 & 0.00062\\
0.6 & -4.19874 & -4.19806 & -4.19857 & 0.00051\\
\end{tabular} 
\end{ruledtabular} 
\end{table} 

From these results it is immediately clear {\it why} thermal fluctuations alone fail to select a unique ground state 
for any value of $b$ considered in this paper.  The fluctuation entropy per site gained by choosing the cubic 16--sublattice 
state is miserly, for example, 
$\Delta s_{\sf f} = 0.00183$ for $b=0.1$ and  
$\Delta s_{\sf f} = 0.00051$ for $b=0.6$.  
These numbers must be compared with the extensive entropy $S_0/N \approx 0.13$ of the liquid phase, {\it all} 
of which is lost if the system orders.   So for the values of $b$ considered here, thermal fluctuations cannot drive the system to order. 

However it is amusing to note that the entropy gain $\Delta s_{\sf f}$ {\it increases} as $b$ decreases, scaling 
approximately as $\ln b$, as shown in Table~\ref{tab:s_f}.    This raises the intriguing possibility 
that $b$ acts as a singular perturbation, and that for sufficiently small $b$, fluctuations might overcome 
the extensive entropy $S_0/N \approx 0.13$ of the dimer manifold, driving the system order --- even 
though it is {\it disordered} for $b =0$.     
Such an order-from disorder effect would presumably favor the cubic 16--sublattice {\sf R}-state, which is also believed 
to be selected by quantum fluctuations at $T=0$~\cite{bergman05,bergman06,sikora09}.   
However in the present model, it would occur only for vanishingly low temperatures, and would therefore be extremely 
difficult to access in simulation.  This question remains for future study.

\begin{figure}[t]
  \centering
  \includegraphics[width=8.5truecm]{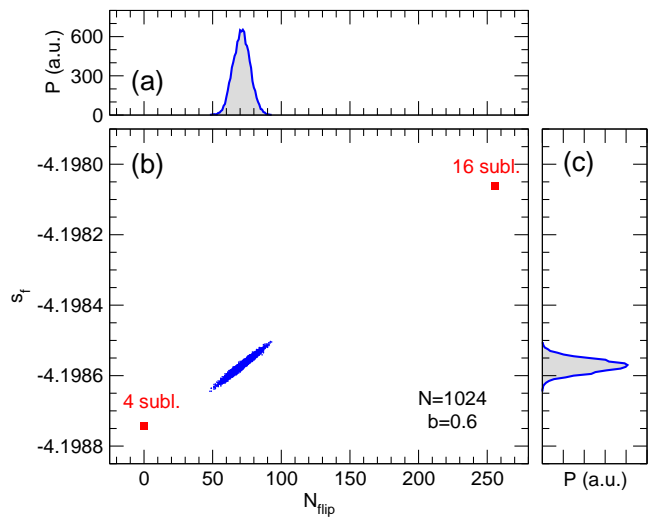}
  \caption{(Color online) (a) Probability distribution of the flippable hexagons within the dimer manifold.   
  (b) The fluctuation entropy per site $s_{\sf f}$ [Eq.~(\ref{eq:s_f})]
  as a function of the number of ``flippable'' hexagons.   
  The lower bound  $s_{\sf f} = -4.19874$ is set by the four--sublattice state shown in \protect\mbox{Fig.~\ref{fig:liquid-solid2}(b),}    which has no flippable hexagons.
 The upper bound $s_{\sf f} = -4.19806$ is set by the 16--sublattice state with the maximum number of flippable hexagons [see \protect\mbox{Fig.~\ref{fig:liquid-solid2}(c)}].  The blue dots represent a sample of 10000 random configurations.
  (c) Probability distribution of the fluctuation entropy per site $s_{\sf f}$ 
  within the dimer manifold.   All results are for a cluster of $N=1024$ sites with $b=0.6$.
\label{fig:lnDetM_Nflip}}
\end{figure}

The result above explains why the system does not order at finite temperature, but not {\it why} the fluctuation entropy favors 
the 16--sublattice state? We can answer this question by looking at the distribution of the fluctuation entropies $s_{\sf f}$ 
within the dimer manifold of $uuud$ states. 
Figure~\ref{fig:lnDetM_Nflip} shows the distribution for $b=0.6$. 
The $uuud$ states can be broken up into classes of states with a different net flux of an effective magnetic (or, equivalently, electric) 
field~\cite{izakov04, hermele04, henley05,bergman05,bergman06, sikora09}.   This flux is conserved by all cyclic exchanges of spins on 
loops of alternating $u$ and $d$ spins, motivating a loop expansion of the fluctuation contribution to the free 
energy~\cite{hizi05,hizi06,henley06,hizi07,bergman07PRB}. The leading term in such an expansion counts the number 
$N_{\sf flip}$ of six--site hexagonal rings in a ``flippable'' \mbox{$u$--$d$--$u$--$d$--$u$--$d$} configuration.   

In Fig.~\ref{fig:lnDetM_Nflip}(b) we plot the fluctuation entropy $s_{\sf f}$ as a function of $N_{\sf flip}$.  
The highest (lowest) values are achieved for the 16--sublattice (four--sublattice) ordered states with the most (least) flippable hexagons. The fact that randomly generated $uuud$ states lie extremely close to the line connecting
these two states suggests that loops of more than six sites contribute little to the fluctuation entropy. 

\begin{figure}[t]
  \centering
  \includegraphics[width=8.5truecm]{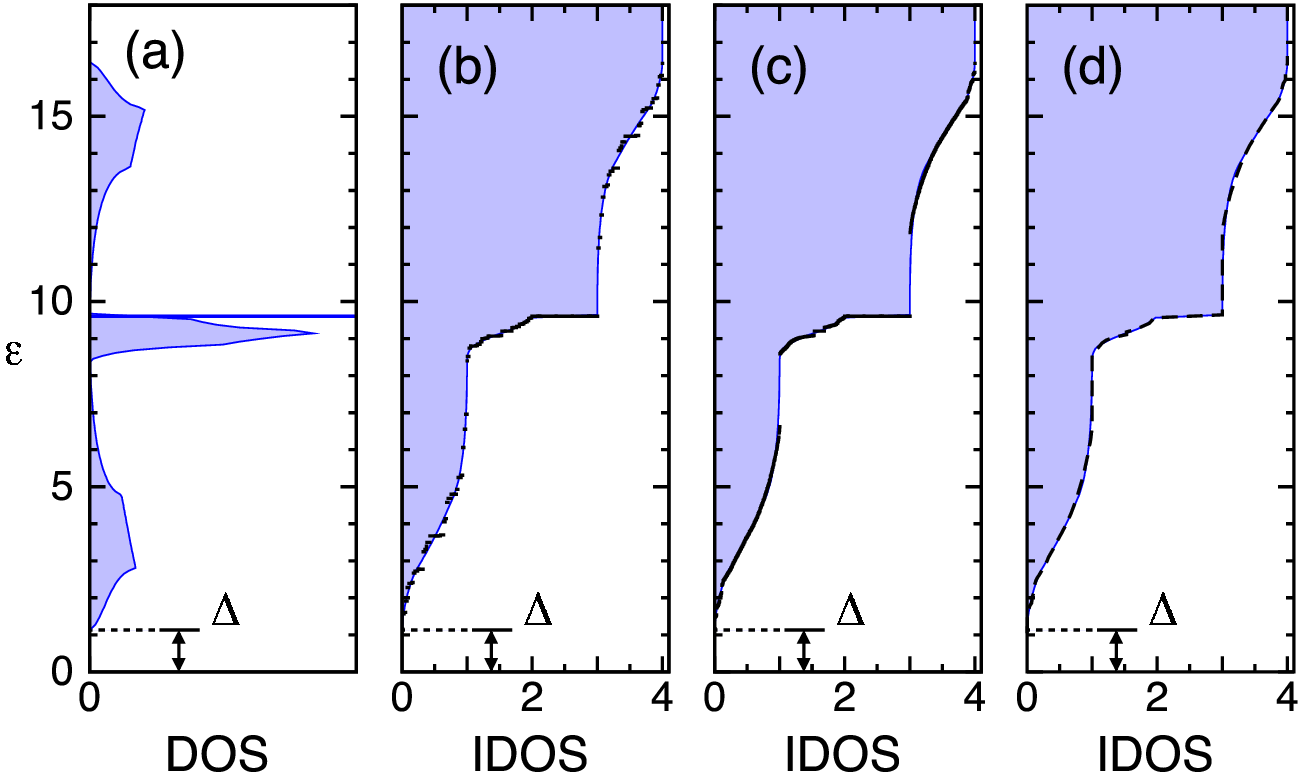}
  \caption{(Color online)  (a) Density of states (DOS) for eigenvalues of ${\cal M}$ 
  for four--sublattice $uuud$ state in thermodynamic limit, showing 
  finite gap $\Delta$ [Eq.~(\ref{eq:Delta})] and flat band at $\varepsilon = 16b$.    
  %For a 1024 spin cluster : 
  Integrated DOS (IDOS)
  (b) of four--sublattice $uuud$ state (points),
  (c) of a typical disordered $uuud$ state (points), 
  and (d) averaged within disordered $uuud$ states (dashed line).  
  In all cases   $J_3=0$, $h=4$, and $b=0.6$, and a cluster size is $N=1024$.
  The integrated DOS corresponding to (a) is shown in
  shading on (b)--(d) for comparison.
\label{fig:DOS}}
\end{figure}

The remaining question is why the overall difference in fluctuation entropy between different $uuud$ states is so small?  
We can express the fluctuation entropy in terms of the $2N$ eigenvalues $\{\varepsilon_n \}$ of ${\cal M}$ as
\begin{eqnarray}
s_{\sf f}  = - \frac{1}{2N} \sum_n \ln \varepsilon_n .
\end{eqnarray}
The eigenvalue spectrum $\{\varepsilon_n \}$ associated with the simplest $\mathbf{q}=0$ four--sublattice $uuud$ state can easily be calculated 
analytically; working with a four--site unit cell, there are four bands, one of which is nondispersive.   The associated density 
of states (DOS) for $b=0.6$ is shown in Fig.~\ref{fig:DOS}(a), where the flat band appears as a sharp peak at $\varepsilon =  16b + h - 4$.   
In Figs.~\ref{fig:DOS}(b)--(d) we compare the integrated DOS from these four bands  with numerical results for the integrated DOS of a 1024--site cluster.   

The integrated DOS, averaged within the dimer manifold of $uuud$ states [Fig.~\ref{fig:DOS}(d)], is indistinguishable by eye 
from that of the four--sublattice state [Fig.~\ref{fig:DOS}(b)].    The step associated with the flat band survives as a set of $N/4$ localized excitations 
at $\varepsilon =  16b$.   And, critically, the gap 
\begin{equation}
\Delta = 8b + 4 - \sqrt{(8 b + 4)(8 b + 4- h) + h^2} 
\label{eq:Delta}
\end{equation}
to the lowest lying excitation is set by a {\it nodeless} eigenvector, whose components $\delta S^\alpha_i$ depend only on whether 
the spin ${\bf S}_i$ points up or down.   All $uuud$ states can be made formally equivalent to four--sublattice order by renumbering 
the sites in each individual tetrahedron, and the energy of this nodeless excitation is also unchanged by this renumbering of sites.  
It is therefore completely insensitive to whether or not the system is ordered.   From the results it is clear why the thermodynamic 
properties of the plateau liquid state, and in particular the entropy associated with fluctuations about it, are so close to those of the 
ordered plateau state.   

From these results, it is also possible to understand why the numerically determined entropy gain $\Delta s_{\sf f}$ increases 
as $b \to 0$ for $h=4$ (cf. Table~\ref{tab:s_f}).   This singular behavior can be traced back to a band of excitations 
above the spin-wave gap $\Delta \approx 4b$, with bandwidth $\Delta \varepsilon \sim b$, which collapses to become a strict 
set of zero modes for $b \to 0$.   Since zero modes are excluded from the sum which determines $\Delta s_{\sf f}$, while the collapsing
band contributes as $\sim \ln b$, $b$ acts as a singular perturbation, and infinitesimal $b$ may drive the system to order.  
This is despite the fact that it is disordered for $b=0$, and for the relatively large $b$ used in our simulations.

\subsection{vector--multipole phase with local ${\sf T_2}$ symmetry}
\label{vectorphase}

At the upper critical field of the magnetization plateau, the collinear spins of the $uuud$ configurations cant away from the $z$ axis.  This instability occurs at the level of a single tetrahedron (Fig.~\ref{fig:tetrahedron}), where it is continuous.   On a lattice, it is associated with the closing of the gap $\Delta$ [Eq.~(\ref{eq:Delta})] in the excitation spectrum of the plateau liquid.   Because of the special structure of this excitation, discussed above, the gap closes at the same value of $h_c = 4+8b $ for all $uuud$ states, and the transition is once again continuous --- at least for \mbox{$T = 0$}.  
However, since the spin configurations in question are simply 3:1 canted versions of the $uuud$ states, with local ${\sf T_2}$ symmetry, 
all of the entropic arguments presented above for the plateau liquid still hold.    Thermal fluctuations alone cannot restore (canted) N\'eel order, and spin--spin correlations exhibit a power--law decay of $1/r^3$  
for \mbox{$T \to 0$}.

The resulting state does however exhibit long range order in {\it both} the rank--two tensor order parameters ${\bf Q}^{\perp,1}$ and 
${\bf Q}^{\perp,2}$ [Eqs.~(\ref{eq:Qperp2}) and (\ref{eq:Qperp1}), and Table~\ref{tab:quadrupoles}].   The 3:1 canting of the $uuud$ 
spins selects a direction in the $xy$ plane, and the primary order parameter is therefore the lower--symmetry irrep, ${\bf Q}^{\perp,1}$.   The finite value of the nematic order parameter ${\bf Q}^{\perp,2}$ reflects the fact that this canting is coplanar.  Since ${\bf Q}^{\perp,1}$  transforms like a vector under 
rotations about the $z$ axis, we classify this state as a vector--multipole phase with local ${\sf T_2}$ symmetry.    

Within the framework of a Ginzburg--Landau theory, the 
contribution to the free energy from this pair of order parameters is 
\begin{eqnarray}
\label{eq:Fvector}
\mathcal{F} & =&  a_1 |{\bf Q}^{\perp,1}|^2  +  a_2 |{\bf Q}^{\perp,2}|^2 \nonumber \\ 
 && + b_{12} \left( Q^{\perp,2}_1 \left[ (Q^{\perp,1}_1)^2 -  (Q^{\perp,1}_2)^2 \right] 
  + 2 Q^{\perp,2}_2 Q^{\perp,1}_1 Q^{\perp,1}_2 \right) \nonumber\\
 && + c_{11}  |{\bf Q}^{\perp,1}|^4 + 2 c_{12}  |{\bf Q}^{\perp,1}|^2 |{\bf Q}^{\perp,2}|^2 + c_{22}  |{\bf Q}^{\perp,2}|^4 \, , \nonumber\\ 
 &&
\end{eqnarray}
where, following Eqs.~(\ref{eq:Qperp2}) and (\ref{eq:Qperp1}), $Q^{\perp,2}_1= Q^{x^2-y^2}$, $Q^{\perp,2}_2= Q^{xy}$, 
$Q^{\perp,1}_1= Q^{xz}$, and $Q^{\perp,1}_2= Q^{yz}$.

\begin{figure}[t]
  \centering
  \includegraphics[width=7truecm]{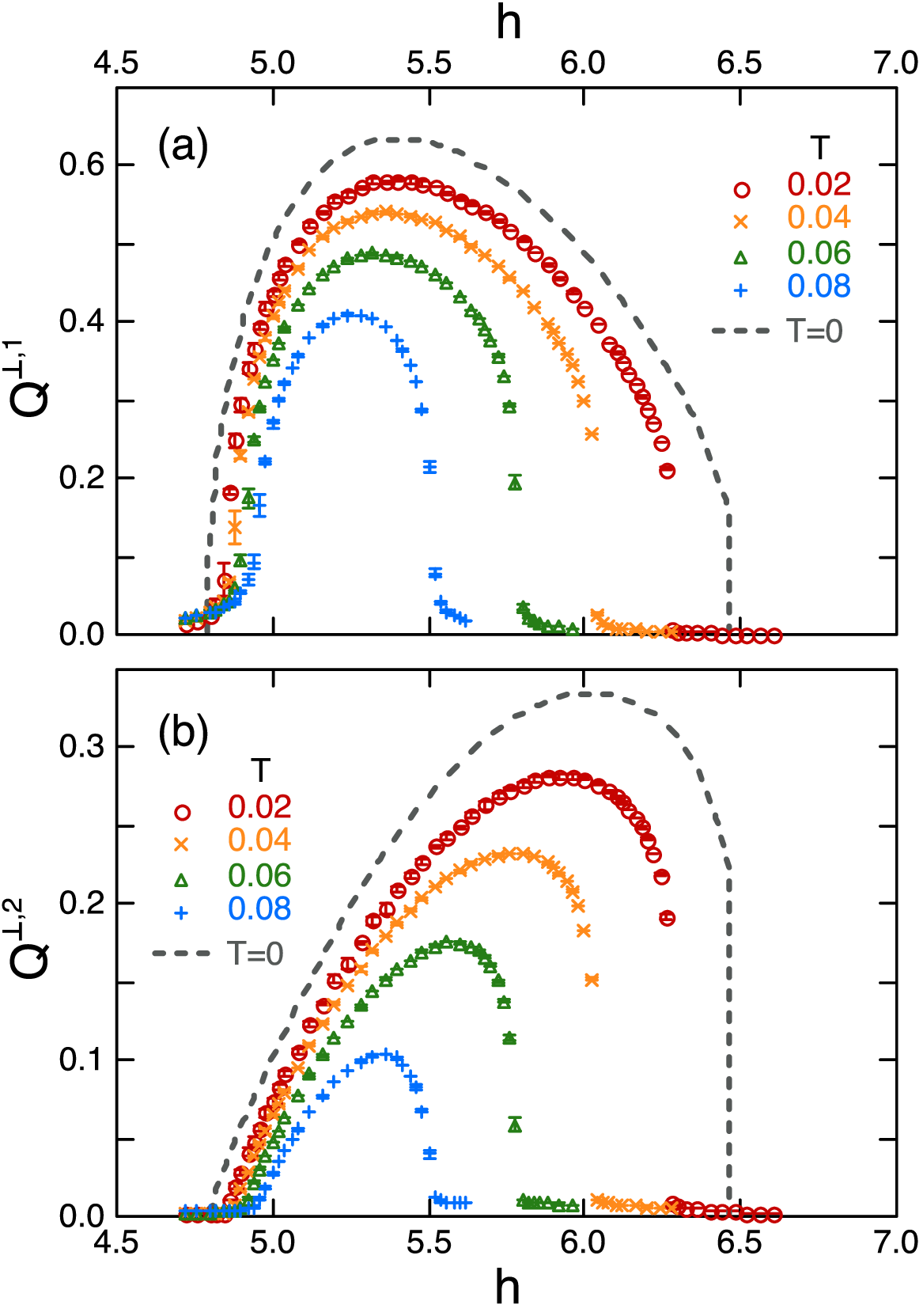}
  \caption{(Color online) Magnetic field dependence of (a) primary order parameter 
  $Q^{\perp,1} = |{\bf Q}^{\perp,1}|$ [Eq.~(\ref{eq:Qperp1})] and (b) 
  secondary order parameter $Q^{\perp,2} = |{\bf Q}^{\perp,2}|$ [Eq.~(\ref{eq:Qperp2})]
  in the 
  vector--multipole phase  for $b=0.1$.   The continuous 
  transition from the  plateau liquid state into the 
  vector--multipole phase at lower magnetic fields, and the direct transition from the 
  vector--multipole phase into the (saturated) paramagnet at higher fields are clearly visible.   
  Dashed lines show behavior at $T=0$ in a   
  single-tetrahedron theory.
  Points show results of MC simulations for the system size $N=8$ from $T=0.02$ to $T=0.08$.
  \label{fig:T2vectorMFT}}
\end{figure}

\begin{figure}[t]
  \centering
  \includegraphics[width=8truecm]{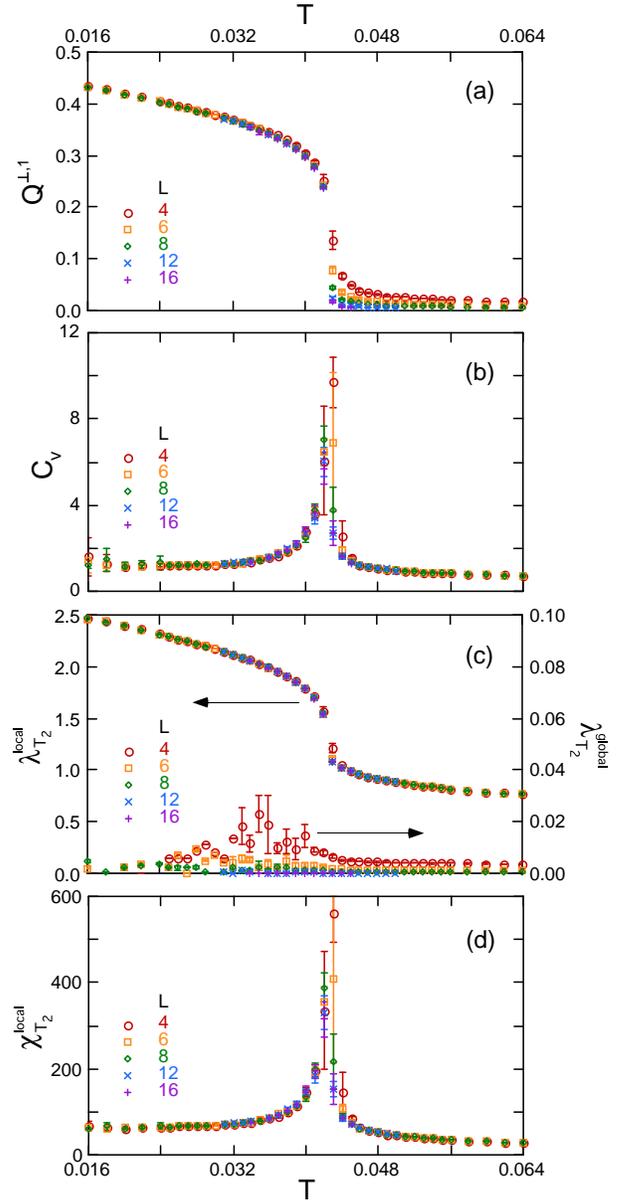}
  \caption{(Color online) Temperature dependence of 
  (a) the primary order parameter $Q^{\perp,1} = |{\bf Q}^{\perp,1}|$ defined by 
  \protect\mbox{Eq.~(\ref{eq:Qperp1})}, 
  showing the onset of the vector--multipole order at $T=T_V \approx 0.042$, 
  (b) heat capacity \protect\mbox{Eq.~(\ref{eq:Cv})}, 
  (c) the related measure of {\it local} 
  correlation $\lambda_{{\sf T_2}}^{\sf local}$ defined by \protect\mbox{Eq.~(\ref{eq:local})}, 
  and the  {\it global} order parameter $\lambda_{{\sf T_2}}^{\sf global}$ 
  defined by \protect\mbox{Eq.~(\ref{eq:global})}, and
  (d) the associated local susceptibility \protect\mbox{Eq.~(\ref{eq:local_chi})}.  
  Simulations were performed for 
  $h=6$, $b=0.1$, in clusters with $L=4$ to $L=16$. 
  \label{fig:vectorT2}}
\end{figure}

Eq.~(\ref{eq:Fvector}) should be contrasted with the form of free energy in the absence of magnetic field, Eq.~(\ref{eqn:nematicF}).    
The cubic invariant $Q^3$ survives as an interaction $b_{12}$ between ${\bf Q}^{\perp,1}$ and ${\bf Q}^{\perp,2}$,    
which transforms like $e^{2 i \phi} e^{-i \phi} e^{-i \phi} \sim 1$ under rotations about the $z$ axis.   This means that components 
of ${\bf Q}^{\perp,2}$ couple linearly to a quadratic combination of the components of ${\bf Q}^{\perp,1}$.  
Because of this, a finite value of the (lower symmetry) order parameter ${\bf Q}^{\perp,1}$, 
immediately induces a finite value of the (higher symmetry) order parameter ${\bf Q}^{\perp,2}$.   

In principle Eq.~(\ref{eq:Fvector}) permits both first and second order phase transitions into the vector--multipole phase from 
disordered (paramagnetic) or pure nematic phases, depending on the sign of the coefficients $c_{11}$, $c_{12}$, and $c_{22}$.  
The full solution for ${\bf Q}^{\perp,1}$ and ${\bf Q}^{\perp,2}$ 
is further complicated by the fact that these order parameters also couple to octupolar spin moments (see Appendix~\ref{sec:octupoles} for details).    
However the relationship between ${\bf Q}^{\perp,1}$ and ${\bf Q}^{\perp,2}$ is clear at the level of a %simple 
single-tetrahedron theory 
(cf. Ref.~[\onlinecite{penc04}]).   

Within the theory for a single, embedded tetrahedron  --- which is {\it exact} for $T=0$ --- the primary order parameter 
$Q^{\perp,1}=|{\bf Q}^{\perp,1}|$ grows as 
\begin{eqnarray}
Q^{\perp,1}(h \gtrsim h_c)| \cong \frac{3}{\sqrt{2 (3- 2b)}}  \sqrt{h-h_c}  \, ,
\end{eqnarray}
while the secondary order parameter $Q^{\perp,2}=|{\bf Q}^{\perp,2}|$ grows more slowly as 
\begin{eqnarray}
Q^{\perp,2}(h \gtrsim h_c) \cong \frac{3}{2(3-2b)}(h-h_c) \, .
\end{eqnarray}
The results of this 
theory for the ${\sf T_2}$ vector--multipole phase are shown by the dashed lines in Fig.~\ref{fig:T2vectorMFT}.    
For the value of $b$ used 
in the present study, the zero 
temperature transition from  ${\sf T_2}$ vector--multipole phase to paramagnet at high field is strongly first order, even at the level 
of a 
single-tetrahedron theory --- cf. Fig.~\ref{fig:tetrahedron} --- and remains so throughout.

The nature of the finite temperature transition from the vector--multipole phase into the paramagnet is 
harder to determine.  However, as shown in Fig.~\ref{fig:T2vectorMFT}, it appears to be first order for all $h > h^* $, 
where $(T^*, h^*) \approx (0.1, 5.2)$ marks the point at which the crossover line $T^*$ joins the boundary of the vector--multipole phase, $T_V$, 
as shown in Fig.~\ref{fig:hTphasediag}(b).   On the basis of our results, we consider that there is a tricritical point at $(T^*, h^*)$ where
the nature of the phase transition into the ${\sf T_2}$ vector--multipole phase changes from continuous to first order.  

In Fig.~\ref{fig:vectorT2} we present MC simulation results for the finite--temperature transition into the vector--multipole
phase for $h =6.0 > h^*$.   The primary order parameter ${\bf Q}^{\perp,1}$ becomes nonzero with a sharp jump at a transition temperature 
$T_V \simeq 0.042$ [Fig.~\ref{fig:vectorT2}(a)].   Both heat capacity and local ${\sf T_2}$ susceptibility show a jump at the transition $T_V$
[Figs.~\ref{fig:vectorT2}(b) and (d)], but no sign of long range order in the bond--order parameter given by Eq.~(\ref{eq:Tdsym})
[Fig.~\ref{fig:vectorT2}(c)].    

The finite temperature transition from the plateau liquid to ${\sf T_2}$ vector--multipole phase for $h < h^*$
deserves special attention, since the plateau liquid exhibits algebraic decay 
of correlations for intermediate distances.  
Monte Carlo simulations suggest that the transition has a continuous character with (approximately) mean--field exponents.
We return to this below in Sec.~\ref{liquid2vq}.

\subsection{Global structure of the $h$--$T$ phase diagram}

Our results for the $h$--$T$ phase diagram of the antiferromagnetic nearest--neighbor 
Heisenberg model with additional biquadratic interactions $b$ [Eq.~(\ref{eq:Hb})] are summarized 
in Fig.~\ref{fig:hTphasediag}(b).   There are two ordered phases, a nematic phase with local ${\sf E}$ symmetry 
and a vector--multipole phase with local ${\sf T_2}$ symmetry, both of which break spin rotational symmetry about the 
direction of the magnetic field.  These are separated by a plateau--liquid state with all the symmetries of 
a paramagnet in magnetic field.

This phase diagram bears a very strong resemblance to that of the corresponding model with 
weak FM third--neighbor interaction $J_3=-0.05$,  which enforces four--sublattice order, as shown 
in Fig.~\ref{fig:hTphasediag}(a) (cf. Ref.~[\onlinecite{motome06}]).  
So far as the topology of the phase diagram is concerned the {\it only} change is the replacement of a line of first 
order phase transitions terminating the four--sublattice plateau state (which breaks lattice symmetries), 
by a crossover in the case of the plateau liquid (which does not).

Throughout this paper, we have argued that preformed local order at the level of a single tetrahedron 
exists in {\it all} of these phases.   Moreover, in the case of the half--magnetization 
plateau, we have seen in Sec.~\ref{plateauliquid} that conventional magnetic order has very little impact on the excitation 
spectrum, and therefore on the thermodynamic properties of the system.

Viewed in this way, the correspondence between the two $h$--$T$ phase diagrams is not 
at all surprising --- the role of secondary interactions like $J_3$ is merely
to select between an infinite set of different ordered ground states.  
Precisely how enforcement of long range order works at finite temperature is a complex and 
very interesting question, to which we provide only a partial answer below.

\section{Thermal transitions between different ordered and disordered states}
\label{deltaJfinite}

\subsection{General context}

None of the phases described above possess conventional magnetic order of the form $\langle {\bf S}_i \rangle \ne 0$.
However they all contain the seeds of such order in the form of well formed local 
orders $\lambda_{\sf E}^{\sf local}$ and $\lambda_{\sf T_2}^{\sf local}$.  Long range order can easily be restored 
by adding additional terms to the Hamiltonian Eq.~(\ref{eq:Hb}).  The simplest possible choice is a FM third--neighbor
interaction $J_3 < 0$ in Eq.~(\ref{eq:HJ3}), leading to four--sublattice order of the form considered in Refs.~[\onlinecite{penc04}] and [\onlinecite{motome06}].   
In what follows we study how FM $|J_3| \ll b$ precipitates an ordered $uuud$ state from the plateau liquid 
for $h=4$, and contrast this with the way in which N\'eel order emerges from the nematic phase with local {\sf E} symmetry for $h=0$. 
We also discuss the continuous transition from plateau liquid to vector--multipole phase 
for $h < h^*$, $T < T^*$.

We study these phase transitions as a function of temperature $T$ which also gives us access 
to the high temperature paramagnetic phase.    This is interesting because, for intermediate distances $r < \xi_c \sim \sqrt{T}$ 
the nematic phases exhibit the algebraic decay of spin correlations characteristic of a Coulomb phase, 
rather than the exponential decay of correlations more usually associated with a paramagnet.   
Transitions between a disordered phase subject to an ice--rule type constraint and a phase with conventional order have 
been discussed for a long time in the context of hydrogen--bonded ferroelectrics~\cite{slater41,lieb67,youngblood80,youngblood81}.   
More recently such questions have arisen again in the context of experiments of many highly frustrated 
magnets~\cite{mirebeau02}, and in the past few years there has been a theoretical effort to understand how
order can emerge from a Coulomb phase in classical dimer~\cite{alet06,misguich08,powell08,charrier08,chen09} and 
spin models~\cite{pickles08,pickles-thesis}.

A strong motivation for this work has been the possibility of observing an unusual continuous phase transitions, including
transitions lying outside the Landau--Ginzburg--Wilson paradigm~\cite{senthil04}.  Indeed the (classical) dimer model on cubic lattice
does exhibit a continuous transition from a Coulomb phase at high temperatures to a simple crystalline ordered phase as a function
of temperature~\cite{alet06}.    This transition has unusual scaling properties~\cite{misguich08}, and does not naively admit a Landau--Ginzburg--Wilson 
description, since the high temperature phase cannot be described using an expansion in terms of the low--temperature order parameter.
A recent very detailed simulation study of a family of three--dimensional dimer models with ordered ground states and high--temperature Coulomb phases
found a rich variety of continuous and discontinuous phase transitions, including double phase transitions where monopole excitations condense 
out of the Coulomb phase to give a conventional paramagnet at intermediate temperatures~\cite{chen09}.    

The present understanding of these phenomena is that the gauge field associated with Coulomb phase is minimally coupled to 
a matter field which condenses in the ordered phase, following an Anderson--Higgs mechanism~\cite{powell08,charrier08,chen09}.
In fact it is also possible to study zero temperature (quantum) phase transitions from Coulomb to ordered phases in three--dimensional {\it quantum} 
dimer models~\cite{moessner03}.   These can in principle be continuous, occurring through the condensation of monopole excitations in the Coulomb 
phase~\cite{bergman06PRB}, but numerical simulations suggest that the transition is first order~\cite{sikora09}.

Less is known about transitions in spin models, but one interesting scenario exists for a continuous transition in an extended 
Heisenberg model on a pyrochlore lattice from a Coulomb phase to a four--sublattice ordered state~\cite{pickles08}.   
This transition is found to be in the same universality class as a uniaxial ferroelectric with dipolar interactions, for which
the upper critical dimension is three~\cite{larkin69}.   
This makes possible to continuous transitions with mean-field exponents (up to log corrections) --- a scenario which 
closely resembles the transition from plateau-liquid into vector-quadrupole phase discussed below.   
Generically, however, transitions from Coulomb liquids into ordered states seem to be first order~\cite{pickles08}, 
a fact which may be explained by interactions between fluctuations of associated gauge field~\cite{pickles-thesis}.   

We conclude by noting that the complex forms of order which can
occur in Heisenberg models on the pyrochlore lattice as a result of the interplay between farther--neighbor interactions 
and thermal fluctuations are also a topic of current interest~\cite{chern08}.  
In finite magnetic field, these lead to a half--magnetization plateau which can be tuned at will between 
different forms of order~\cite{motome09}.   A similar fluctuation driven plateau, but with a uniquely defined 
form of order, is also expected to occur for the edge sharing tetrahedra of the FCC lattice~\cite{zhitomirskyXX}.

We now return to the model in question.

\begin{figure}[t]
    \centering
    \includegraphics[width=.38\textwidth]{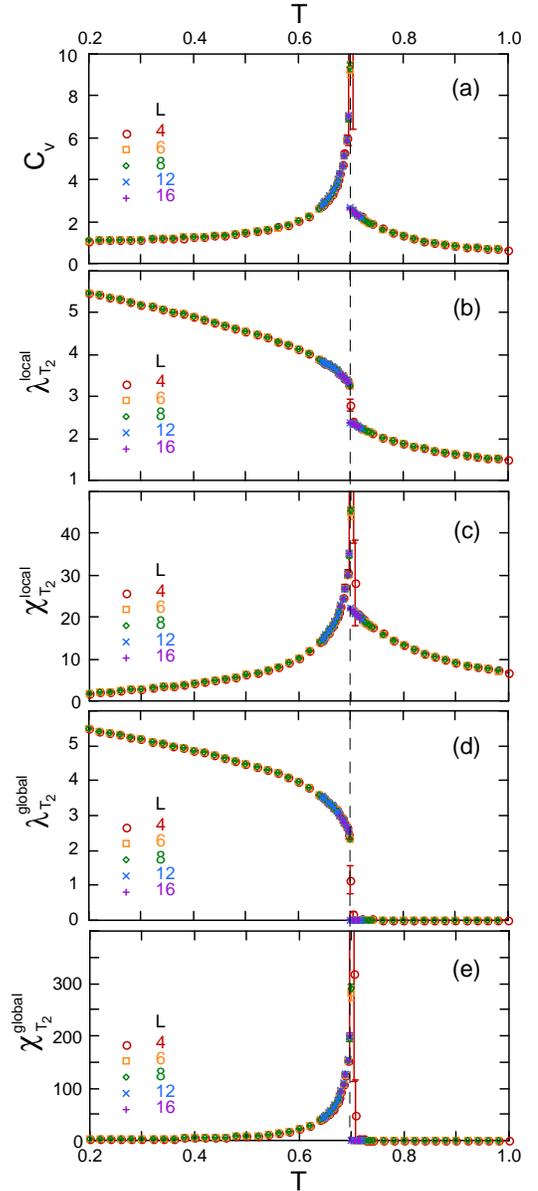}
    \caption{(Color online) Temperature dependence of (a) heat capacity
    defined by \protect\mbox{Eqs.~(\ref{eq:Cv})}, (b) and (c) the related measure of {\it local} 
  correlation  $\lambda_{{\sf T_2}}^{\sf local}$ and its susceptibility defined by \protect\mbox{Eqs.~(\ref{eq:local}) and (\ref{eq:local_chi})}, 
  and (d) and (e)  the  {\it global} order parameter $\lambda_{{\sf T_2}}^{\sf global}$ and its susceptibility defined 
  by \protect\mbox{Eqs.~(\ref{eq:global}) and (\ref{eq:global_chi})} for $b=0.6$, $h=4$ and a range of values of system size.
  The results are for $J_3=-0.06$, showing a single first order transition into the ordered phase at $T_N=0.70(1)$ (indicated by the vertical dashed line).  
}
    \label{fig:comparison1}
\end{figure}

\begin{figure}[t]
    \centering
    \includegraphics[width=.38\textwidth]{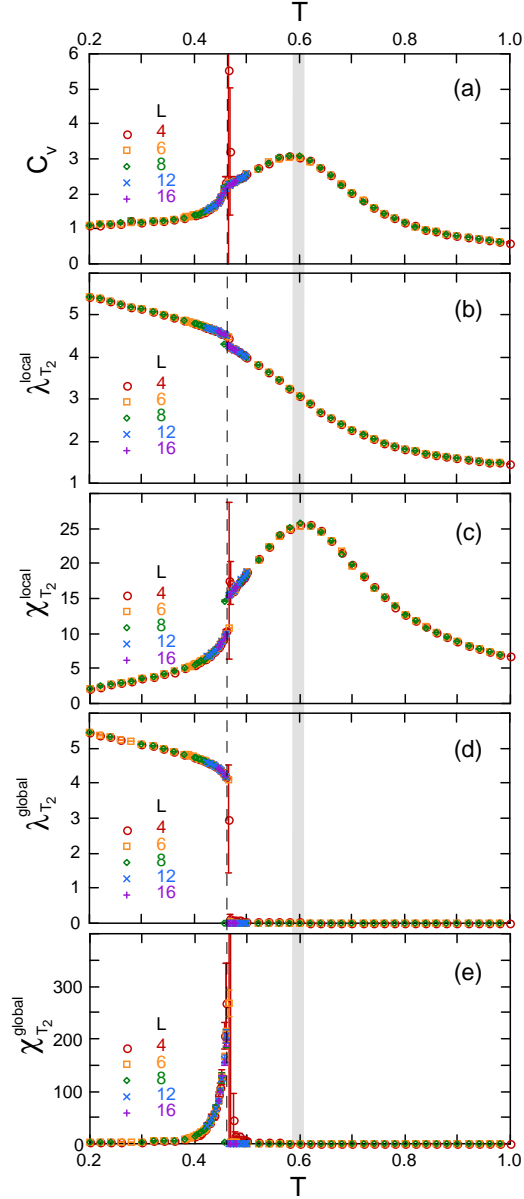}
    \caption{(Color online) The same plots as Fig.~\ref{fig:comparison1} but for \mbox{$J_3=-0.02$}, showing both a crossover in to the plateau liquid for $T^*\approx 0.6$ (bold grey line) and 
  a transition into the ordered phase at $T_N=0.46(1)$ (dashed line).
}
    \label{fig:comparison2}
\end{figure}

\begin{figure}[t]
    \centering
    \includegraphics[width=7truecm]{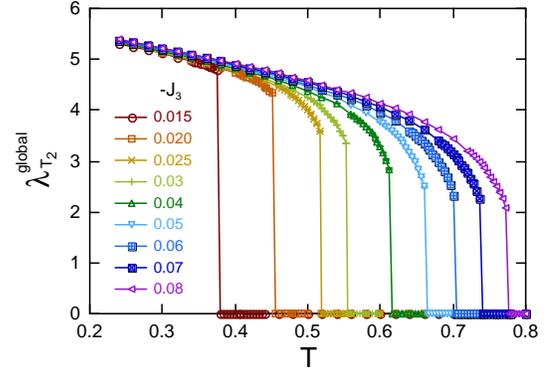}
    \caption{(Color online) Temperature dependence of the order parameter 
    $\lambda_{{\sf T_2}}^{\sf global}$ for four-sublattice $uuud$ order [as defined by 
    \protect\mbox{Eq.~(\ref{eq:global})}], for a range of values of $J_3 < 0$.    
    The transition temperature becomes smaller as $|J_3| \to 0$.  
    At the same time the transition becomes more strongly first-order.  
    All results are for $h=4$ and $b=0.6$, in a cluster with $L=8$.  
    The lines are guides for the eye.
}
    \label{fig:LT_2_J3dep}
\end{figure}

\begin{figure}[t]
  \centering
  \includegraphics[width=7.5truecm]{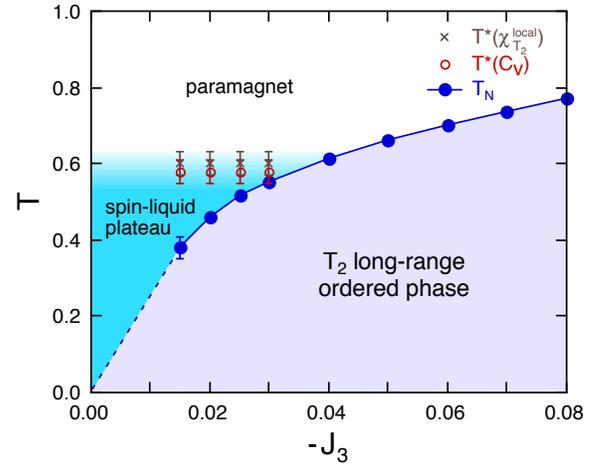}
  \caption{(Color online) 
  Phase diagram for the classical Heisenberg antiferromagnet on a pyrochlore 
  lattice in applied magnetic field $h=4$, with additional
  biquadratic interactions $b=0.6$ [Eq.~(\protect\ref{eq:Hb})].
  The transition temperature $T_N$ associated with the gapped, ordered, half--magnetization plateau state
  vanishes as the strength of ferromagnetic third--neighbor interactions $J_3 \to 0$, as determined by Monte Carlo simulation.   
  A state exhibiting a half--magnetization plateau but no long--range magnetic order 
  exists above $T_N$ up to a crossover temperature $T^*$.    Estimates of the crossover temperature 
  are taken from peaks in the local susceptibility and heat capacity.  
\label{fig:J3}}
\end{figure}

\begin{figure}[t]
    \centering
    \includegraphics[width=.38\textwidth]{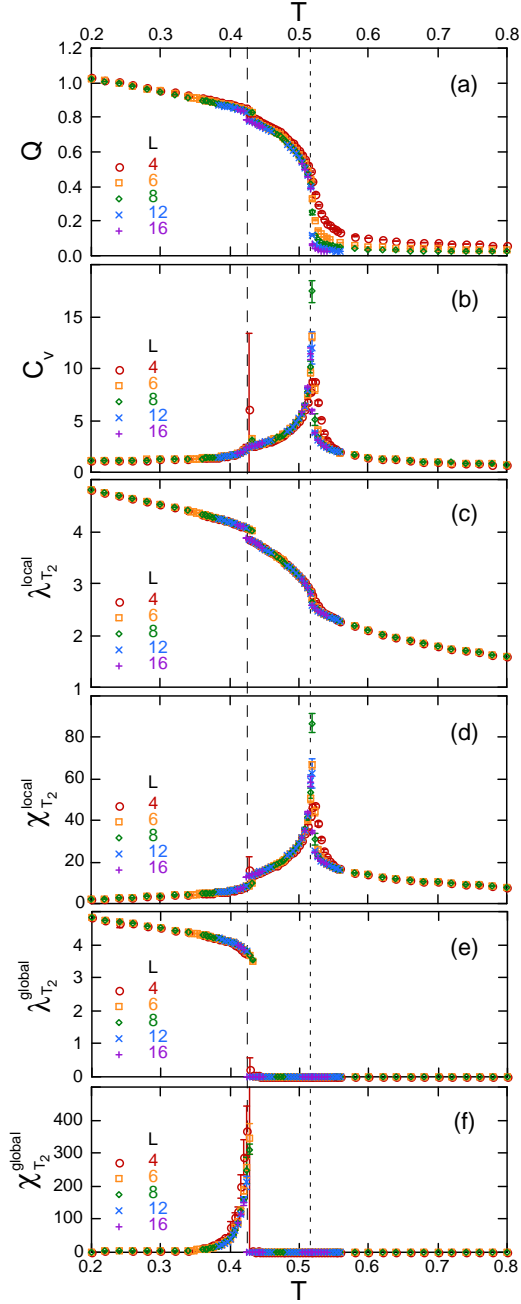}
    \caption{(Color online) Temperature dependence of 
    (a) nematic order parameter [Eq.~(\ref{eq:Q})],
    (b) heat capacity [Eq.~(\ref{eq:Cv})], (b) and (c) the related measure of {\it local} 
  correlation  $\lambda_{{\sf E}}^{\sf local}$ and its susceptibility 
  [Eqs.~(\ref{eq:local}) and (\ref{eq:local_chi})],
  and (d) and (e)  the  {\it global} order parameter $\lambda_{{\sf E}}^{\sf global}$ and its susceptibility [Eqs.~(\ref{eq:global}) and (\ref{eq:global_chi})]
  for $b=0.6$, $h=0$ and a range of values of system size.
  The results are for $J_3=-0.02$, showing both a first order transition into the nematic phase at $T_Q=0.518(6)$ (indicated by the vertical dotted line) and
  a transition into the ordered phase at $T_N=0.42(2)$ (dashed line).}
    \label{fig:comparison3}
\end{figure}

\begin{figure}[t]
  \centering
  \includegraphics[width=7.5truecm]{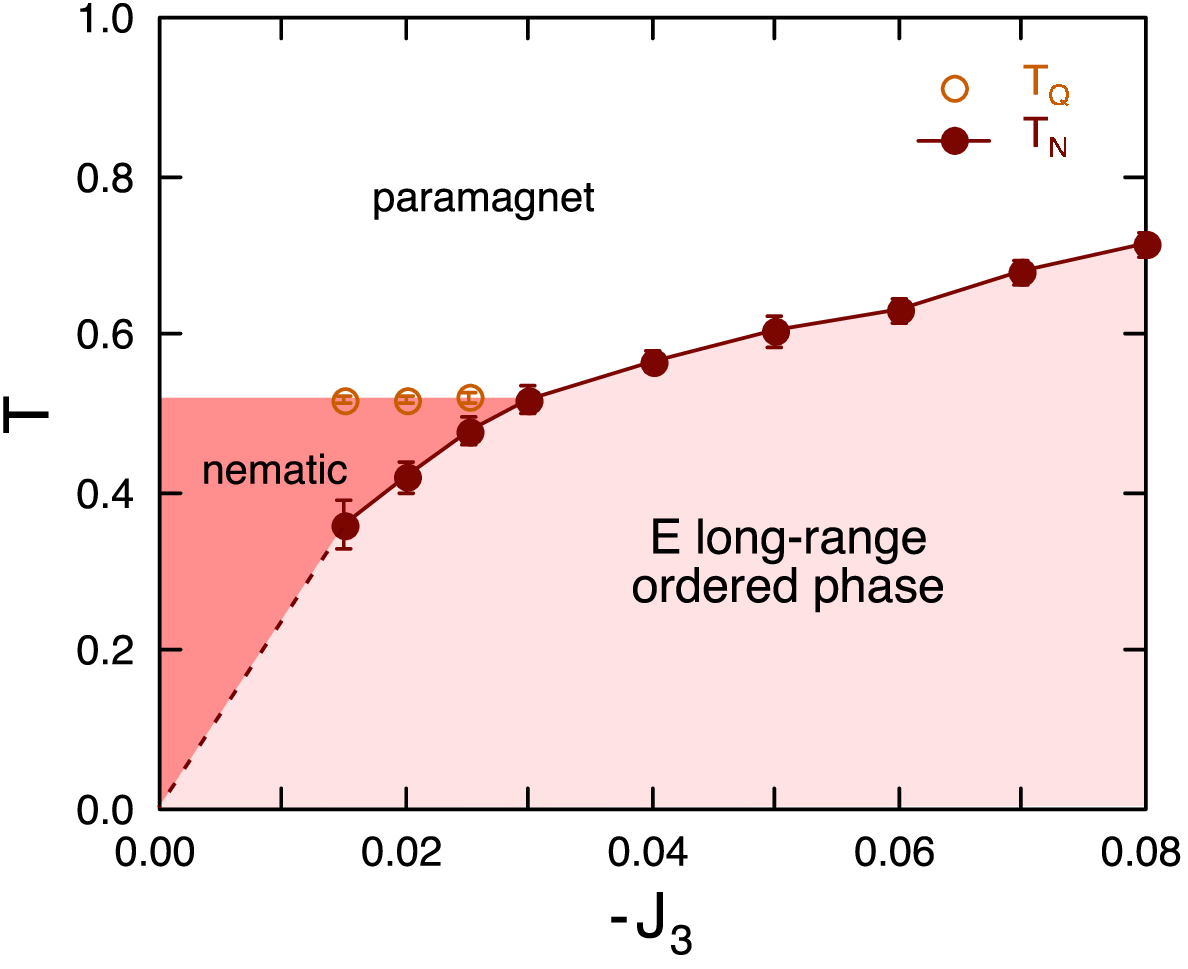}
  \caption{(Color online) 
  Phase diagram for the classical Heisenberg antiferromagnet on a pyrochlore 
  lattice, Eq.~(\protect\ref{eq:Hb}), in applied magnetic field $h=0$, with additional
  biquadratic interactions $b=0.6$.
  The transition temperature $T_N$ associated with the {\sf E}--symmetry long--range order
  vanishes as the strength of ferromagnetic third--neighbor interactions $J_3 \to 0$, as determined by Monte Carlo simulation.   
  A  phase exhibiting nematic order exists above $T_N$ up to a  $T_Q \sim b$. 
\label{fig:J3-2}}
\end{figure}

\subsection{Transition from plateau--liquid to ordered $uuud$ state}

For $h \simeq 4$,  $T \lesssim b$, Eq.~(\ref{eq:Hb}) exhibits the plateau--liquid state described in Sec.~\ref{plateauliquid}
[cf. Fig.~\ref{fig:hTphasediag}(b)].    Inclusion of a FM third--neighbor interaction $J_3$ [Eq.~(\ref{eq:HJ3})] causes it 
to order at low temperatures.   We consider first the conventional limit where both $|J_3|$ and $b$ are  ``large'', 
choosing parameters  $J_3=-0.06$ and $b=0.6$. 
In this case there is strongly first order transition from paramagnet 
to four--sublattice plateau state for $T_N \approx 0.70$.    
This can be seen very clearly in simulation results for the heat capacity and the order parameter $\lambda_{\sf T_2}^{\sf global}$, 
and its susceptibility 
$\chi_{\sf T_2}^{\sf global}$, presented in Fig.~\ref{fig:comparison1}.   
If we now decrease $|J_3|$, the transition temperature $T_N$ must also decrease, and for sufficiently small
$|J_3|$ it will become smaller than the crossover temperature $T^* \approx b$ associated with the plateau liquid.

In this case, there are anomalies in thermodynamic quantities at two distinct temperatures as demonstrated in Fig.~\ref{fig:comparison2}.   
There is a broad maximum in $\chi_{\sf T_2}^{\sf local}$ at $T^*\approx 0.6$ [Fig.~\ref{fig:comparison2}(c)], signaling the onset of 
the plateau liquid state, accompanied by a broad peak in the heat capacity $C_v$ at a slightly lower temperature [Fig.~\ref{fig:comparison2}(a)].
And, at $T=T_N \approx 0.46 < T^*$, there is a small jump in $C_v$, accompanied by a clear singularity in global
order parameter susceptibility $\chi_{\sf T_2}^{\sf global}$ [Fig.~\ref{fig:comparison2}(e)].   
While there is no true phase transition at $T^*$, it is clear that the bulk of the entropy of the paramagnet is lost 
in the smooth crossover into the plateau liquid, and not in the first order transition into the ordered phase.

So what happens for $J_3 \to 0$?   Unfortunately this question is hard to answer by Monte Carlo 
simulation, as the massive degeneracy of the $uuud$ states translates into many competing local 
minima in the free energy.    However $T_N$ is strictly zero for $J_3=0$, and there are two obvious scenarios
for how this can be achieved.

The first is that the first--order transition into the ordered phase becomes weaker as $T_N \to 0$, terminating
in a critical end point for $J_3=0$, $T_N=0$.    This end point would in fact be {\it multicritical}, since many different
ordered $uuud$ states can be formed out of the dimer manifold for different choice of long range interactions.
Within this scenario, the power--law correlations between spins in the plateau liquid for $T \to 0$ could be viewed
as evidence of critical fluctuations.  
The second scenario is that the first order transition into the ordered phase persists down to $T_N=0$.
Since an infinite number of different ordered phases branch out from the point $J_3=0$, $T_N=0$, it 
can probably best be termed {\it ``multifurcative''}.   

First--order phase transitions between different ordered phases 
with an infinite degeneracy at the transition occur in a number of models.   
Such phase transitions are first order, in the sense that neither ordered parameter collapses
approaching the critical point.    However they also exhibit one of the characteristic features of a second 
order transition, namely a soft excitation or set of soft excitations connecting the different ordered 
ground states.

As far as we can tell from our present results, it seems most likely that the classical pyrochlore AF
with biquadratic interactions exists at a multifurcative point in parameter space, with an infinite ground--state 
degeneracy, {\it not} at a critical end point.   
As shown in Fig.~\ref{fig:LT_2_J3dep}, 
the low temperature value of the 
order parameter is broadly independent of $T_N$.   This implies that the phase transition in fact becomes 
{\it more strongly} first order at $T_N \to 0$, and appears to rule out a (multi)critical end point.
Our collected simulation results for $J_3 \to 0$  are summarized 
in the form of the phase diagram in Fig.~\ref{fig:J3}.   

It is amusing to note that this phase diagram bears a superficial resemblance to the phenomenology 
of a second order (quantum) critical point --- a transition temperature which collapses to a special 
point with algebraic decay of correlation functions, which in turn controls a broad region of the phase 
diagram up to a characteristic crossover temperature $T^*$.   All of this despite the fact that the only 
phase transition present is first order, which means that the length scale associated with fluctuations 
remains finite.   Some of the generic features seen in our model --- power law decay of correlations over a large, 
but finite, length scale --- have been previously discussed in the context of other models with strong local constraints, 
where they were dubbed ``high temperature criticality''~\cite{castelnovo06}.

The transition from a critical ``Coulombic'' phase described by a $U(1)$ gauge theory into a simple ordered 
state as a function of temperature can be studied much more cleanly in the (classical) dimer model on cubic lattice, 
where the constraint enforcing the dimer manifold is infinite.   In this case, the phase transition is continuous, 
and exhibits interesting and unusual scaling properties~\cite{alet06,misguich08}.     We have made 
a preliminary study of the ``stiffness'' $K$ associated with fluctuations in a $U(1)$ gauge theory
for temperatures spanning the paramagnet and plateau liquid phases in our model~(see Fig.~\ref{fig:J3}), 
but find no clear evidence of a phase transition.    However the way in which the dimer and loop manifolds 
break down at finite temperature in a model with a finite constraint is an interesting problem, and one which 
deserves further study.  
We note in passing that interesting, related, problems arising the context of quantum loop models~\cite{troyer08}.

\subsection{Transitions from paramagnet to {\sf E}--symmetry nematic phase and N\'eel ordered state}

It is interesting to contrast the finite temperature phase transitions associated with the 
plateau states for $h \approx 4$, with the transitions into {\sf E}--symmetry N\'eel and nematic ordered
states for $h=0$. 
Once again, for ``large'' $|J_3|$ there is a strongly first order transition 
from the paramagnet into the N\'eel phase at a unique temperature $T_N$.   Meanwhile, for  ``small'' $|J_3| \ll  b$, there 
is double transition, first from paramagnet to {\sf E}--symmetry nematic phase at $T_Q \sim b$, and then 
into the four--sublattice N\'eel order at a much lower temperature $T_N$, 
as demonstrated in Fig.~\ref{fig:comparison3}.   
Within the limits of our simulation, both of these transitions appear to be first order in character~\footnote{We observe substantial hystereses and very slow relaxation process in MC calculations in the first order transitions at $h=0.0$ $b=0.6$; We here adopt mixed initial configurations in which a half of the system is nematic ordered and the rest half is paramagnetic disordered.}.   
The results for varying $J_3$ are summarized in the phase diagram in Fig.~\ref{fig:J3-2}.  
The first order transition from paramagnet to nematic phase at $T_Q$ at $h=0$ should be compared with the crossover 
from paramagnet to plateau liquid $T^* \sim b$ observed for $h=4$ in Fig~\ref{fig:J3}.  

\subsection{Transition from plateau--liquid to vector--multipole phase}
\label{liquid2vq}

\begin{figure}[t]
  \centering
  \includegraphics[width=7truecm]{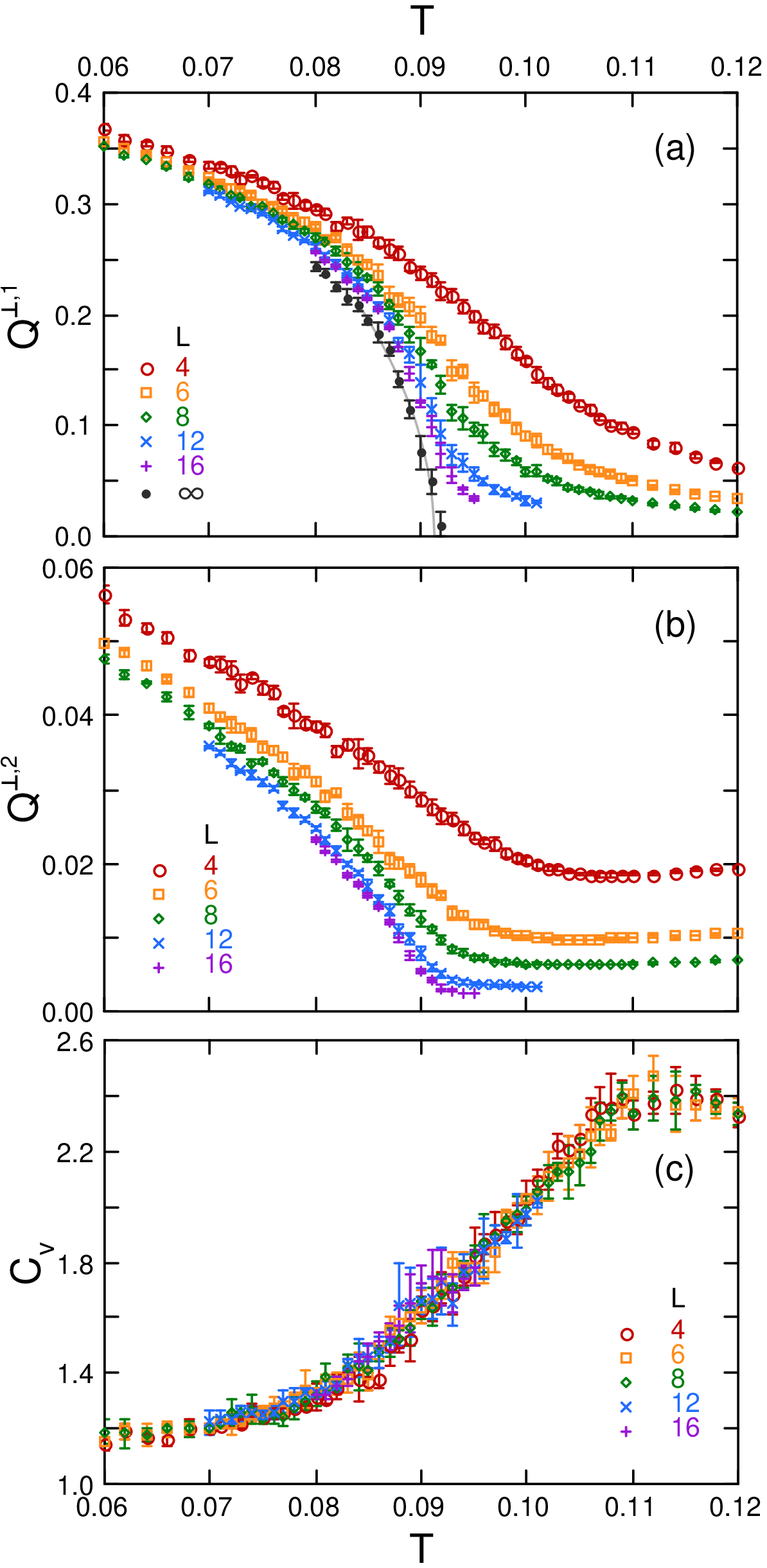}
  \caption{(Color online) 
  (a) Temperature dependence of the primary order parameter $Q^{\perp,1}$ in the 
  vector--multipole phase.  In the limit $L \to \infty$ these results extrapolate to a mean field--like
  behavior $Q^{\perp,1} \sim \sqrt{T_V-T}$ with $T_V \simeq 0.091$ (solid black points and grey line).
  (b) The secondary order parameter $Q^{\perp,2}$ also takes on a finite value for 
  $T < T_V$, but grows more slowly at the transition.    
  (c) Heat capacity, showing no measurable singularity at $T=T_V$.  
  All data are for $h=5$, $b=0.1$, $J_3=0$, with system sizes ranging from $L=4$ to $L=16$.  
  \label{fig:MFtransition}}
\end{figure}

Perhaps the most interesting of the finite temperature transitions observed in our model is the one
from plateau--liquid to vector--multipole phase, already described in Sec.~\ref{vectorphase}.   
At one level this is the most exotic phase transition we study --- a continuous phase transition 
from a ``Coulombic'' state with algebraic decay of correlation functions (the plateau liquid) to a 
phase with long--range multipolar order (the vector--multipole state).  
But at the same time it has the simplest phenomenology of any of the phase transitions in this paper, 
with the order parameter exhibiting a simple mean--field like behavior 
$Q^{\perp,1}(T) \sim \sqrt{T_V-T}$ with $T_V \approx 0.09$, as shown in Fig.~\ref{fig:MFtransition}(a). 
The secondary order parameter $Q^{\perp,2}$ grows more slowly as expected [Fig.~\ref{fig:MFtransition}(b)].
The heat capacity does not show a noticeable singularity 
at $T=T_V$ in Fig.~\ref{fig:MFtransition}(c), 
which is also consistent with the mean--field behavior $C_v \sim (T-T_V)^\alpha$ with $\alpha = 0$.
(The broad peak at $T \sim 0.112$ again corresponds to the crossover temperature $T^*$ 
for the plateau-liquid state.)

At a qualitative level, and in the spirit of this paper, it is easy to see how a continuous transition can arise between these two states.   
Both are built of tetrahedra with a local ${\sf T_2}$ character, with three ``up'' and one ``down'' spin, 
joined at the corners.   Both states will exhibit algebraic decay spin correlations at low temperatures, as a 
result of the infinite number of ways that these tetrahedra can be assembled to form a pyrochlore lattice.
The only difference is that three ``up'' and one ``down'' spins are canted in the vector--multipole phase, 
giving a finite value of $Q^{\perp,1}$ and $Q^{\perp,2}$
[Figs.~\ref{fig:MFtransition}(a) and (b)].   
As long as this canting can interpolate smoothly to 
zero in the collinear $uuud$ state, the transition will be continuous.   And at the level of a Ginzburg--Landau theory, 
nothing prevents this from happening --- cf. Eq.~(\ref{eq:Fvector}).  

However in principle it should also be possible to transcribe each of these phases in terms of the more sophisticated ``solenoidal  field'' theory
used to describe N\'eel order in a spin model with a high temperature Coulomb phase (cf.~Ref.~[\onlinecite{pickles08}]).    To the best 
of our knowledge, nobody has yet attempted to extend the gauge--field description of a Heisenberg type 
spin model to treat multipolar order.   But it is interesting to note that the transition from Coulombic phase to simple N\'eel order
was found to be continuous, and in a universality class with upper critical dimension three, i.e., one 
where the  critical behavior is mean--field like, up to log corrections~\cite{pickles08,larkin69}.

%%%%%%%%%%%%%%%%%%%%%%%%%%%%%%%%%

\section{Summary and conclusions}
\label{conclusions}

We have studied the ordered and disordered phases of the classical, bilinear--biquadratic
Heisenberg model on the pyrochlore lattice at finite temperature and in applied magnetic field.
We find a rich collection of unconventional states --- nematic and vector--multipole phases
with distinct and different local symmetries, separated by a half--magnetization plateau with spin--liquid character.
All of these phases show an underlying ``Coulombic'' character with algebraic decay of spin correlation functions 
over distances $r \lesssim \xi_c \sim 1/\sqrt{T}$.  
Interestingly, the transition from plateau--liquid to vector--multipole phase is continuous, 
and appears to be well--described by mean field theory.
 
While this behavior is undeniably exotic, all of these states can be understood --- and even anticipated --- from 
a proper understanding of the geometry of the pyrochlore lattice, and the properties of a single tetrahedron.
Strong {\it local} fluctuations of N\'eel order are present in all of these phases, 
and the zero temperature phase diagram can be understood simply from the ``self assembly'' of these 
ordered tetrahedra into complex states with higher symmetry.

It is therefore unsurprising that conventional N\'eel order (with four--sublattice structure) is 
immediately restored by the introduction of a ferromagnetic third--neighbor coupling $J_3$.   
However for small $|J_3|$,  the unconventional states survive 
above the N\'eel transition temperature $T_N$.
In particular, the spin--liquid plateau survives above $T_N$, up to a crossover temperature 
$T^* \approx b$.   The transition between liquid and ordered plateaux is first order in nature, and 
remains so for $T_N \to 0$.   For small $|J_3|$, the system also exhibits a first--order transition between the 
N\'eel and nematic phases, in addition to the first--order transition from the nematic phase into high-temperature 
paramagnet.

So far as experiment is concerned, our main finding is that the physics of a pyrochlore antiferromagnet 
in magnetic field can be largely determined by the properties of a single tetrahedron.  
In the simple models which we have considered it is possible to tune between states with entirely different point 
group symmetries at will, simply by changing the form of (weak) long range interactions present.
This is an oversimplification, in the sense that magnetostriction in real systems is likely single out
a particular phonon (or family of phonons) with definite symmetry, which will then drive the system 
towards collinearity ($b$, in our model) {\it and} select the low--temperature ordering pattern 
(long range interactions, e.g., $J_3$, in our model).   However, as long as there is a strong coupling 
to phonons within individual tetrahedra, the form of the magnetization plateau and associated phases 
may be largely independent of these (system dependent) details.

%%%%%%%%%%%%%%%%%%%%%%%%%%%%%%%%%%%%%%%%%%%

\begin{acknowledgments}
We are pleased to acknowledge stimulating discussions with 
F.~Alet,
J.~Chalker, 
G.~Kriza,
G.~Misguich, 
T.~Momoi, 
H.~Shiba, 
H.~Takagi, 
O.~Tchernyshyov, 
S.~Trebst, 
H.~Tsunetsugu, and 
H.~Ueda.  
We are particularly indebted to R.~Moessner and M.~E.~Zhitomirsky 
for valuable comments about multicritical points and the classification of multipolar order.

This work was supported under 
EPSRC Grants No. EP/C539974/1 and EP/G031460/1, and SFB 463 of the DFG (NS);
Hungarian OTKA T049607 and K62280 (KP); Grant--in--Aid for Scientific Research
No. 16GS50219, 17740244, and 19052008 from MEXT, Japan;
Global COE Program ``the Physical Sciences Frontier", MEXT, Japan, 
and Next Generation Super Computing Project, Nanoscience Program (YM).
Part of this work was done while KP and YM were visitors at KITP Santa Barbara. 
KP and NS also acknowledge the hospitality of MPI--PKS Dresden, where
a part of this work was completed.

\end{acknowledgments}

%%%%%%%%%%%%%%%%%%%%%%%%%%%%%%%%%%%%%%%%%%%

\appendix
 
\section{Classification of symmetry breaking at the level of a single site}
\label{symmetry}

In order to identify the different possible forms of magnetic order which can survive where conventional N\'eel order 
breaks down, it is helpful to classify the different forms of symmetry breaking which exist at the level of a single site.   
This analysis is in the spirit of the detailed classification for the nematics in liquid crystals undertaken 
in Ref.~[\onlinecite{lubensky02}], and motivates the rank--two and rank--three tensor order parameters introduced in 
Sec.~\ref{sec:nematic_order_parameter} and Appendix~\ref{sec:octupoles}.    
In order to keep contact with quantum spins, which are axial rather than polar vectors, we must keep track of time reversal symmetry.

In Table~\ref{tab:symspin} we show the transformation rules for the spins under selected symmetry operations, including time reversal
$\Theta {\bf S} = - {\bf S}$.  In contrast to the usual polar vectors, inversion leaves the axial vectors invariant -- as a consequence, all the 
usual (reflection, rotation, and inversion) symmetry operation can be represented by an orthogonal matrix belonging to $SO$(3), with 
determinant equal to +1. The role of inversion in the case of polar vectors is taken over by the time reversal operator $\Theta$.  

All of the symmetry operations, extended with the time reversal, can be represented by orthogonal matrices with determinant -1. 
In the Table~\ref{tab:symspin} we also check if the collinear and coplanar states are invariant under those symmetry operations.  
Since we are interested in the symmetry breaking which can occur in the absence of broken translational symmetry, we do not apply 
the symmetry elements to the lattice points (i.e., we treat all the spins as they were at the origin).  We find that the invariant operations of the 2:2 state include a $C_2(z)$ rotation in addition to the symmetry operations of the 3:1 state.

\begin{table}[hb]
\caption{\label{tab:symspin} The transformation of spins under different symmetry operations. $E$ is the identity element, $I$ is the inversion, $\Theta$ is the time reversal operation, $\sigma_{\alpha\beta}$ is a reflection with a mirror plane $\alpha\beta$, and $C_2(\alpha)$ is a two--fold rotation around the $\alpha$ axis.  In the last two columns we indicate if the 2:2 and 3:1 canted states (with magnetic moment along the $z$ axis and spins are in the $xz$ plane) are invariant with respect to the particular operation. }
\begin{ruledtabular} 
\begin{tabular}{crrrcc} 
symmetry elements & $S^x$ & $S^y$ & $S^z$ & 2:2 & 3:1  \\ 
\hline
$E,I$ & $S^x$ & $S^y$ & $S^z$ & yes & yes \\ 
\hline
$\sigma_{yz},C_2(x)$ & $S^x$ & $-S^y$ & $-S^z$ & no & no \\ 
$\sigma_{xz},C_2(y)$ & $-S^x$ & $S^y$ & $-S^z$ & no & no \\ 
$\sigma_{xy},C_2(z)$ & $-S^x$ & $-S^y$ & $S^z$ & yes & no \\ 
\hline
$\Theta\sigma_{yz},\Theta C_2(x)$ & $-S^x$ & $S^y$ & $S^z$ & yes & no \\ 
$\Theta\sigma_{xz},\Theta C_2(y)$ & $S^x$ & $-S^y$ & $S^z$ & yes & yes \\ 
$\Theta\sigma_{xy},\Theta C_2(z)$ & $S^x$ & $S^y$ & $-S^z$ & no & no \\ 
\hline
$\Theta,\Theta I$ & $-S^x$ & $-S^y$ & $-S^z$ & no & no \\ 
\end{tabular} 
\end{ruledtabular} 
\end{table} 

In Table~\ref{tab:symcanted2} we show the symmetry group of each of the spin states.  In order to facilitate comparison with Ref.~[\onlinecite{lubensky02}], 
we also show the symmetry group of the states if the spins were polar vectors.  We can see that as the time reversal does not play a role for the 2:2 collinear 
state, its symmetry group being the grey--group  $D_{\infty h} + \Theta D_{\infty h}$.   The magnetic (the 3:1 collinear and both canted states) states have a 
magnetic point group as a symmetry group. 
 
When studying which symmetry group is broken for the magnetic states, we need to note that the external magnetic fields lowers the $O(3)$ symmetry of the space to $C_{\infty}\times \{E,I\}+\Theta \sigma_v C_{\infty}\times \{E,I\}$, where the axis of the $C_{\infty}$ is parallel to the magnetic field, and $\sigma_v$ is a reflection to a plane that includes the axis of the magnetic field. The symmetry of the space with magnetic field is actually identical to the symmetry of the 3:1 collinear state.   Thus, within a Ginzburg--Landau framework we do not expect a continuous (second order) phase transition between the $T=0$ liquid plateau and the high temperature disordered phase.  
The $\mathbb{Z}_2$ lowered symmetry of the 3:1 state with respect to 2:2 canted state is manifested in the $\mathbb{Z}_2$ symmetry lowering of the vector to the nematic phase. 

\begin{table*}[b]
\caption{\label{tab:symcanted2} 
The symmetry of the different configurations, treating the arrows as polar vectors, or as axial vectors with and without inclusion of the time reversal symmetry. In the last column we show the broken symmetry (we assume no magnetic field in the case of the 2:2 collinear state and  magnetic field along the $z$ direction for 3:1 collinear and for the two canted states).
The notation is the same as in the Table~\ref{tab:symspin}, with the addition of two elements: $\sigma_v$ is a reflection to a plane perpendicular to the $C_{\infty}$ axis  ($\sigma_{xz}$ is also a $\sigma_v$), while $C'_2$ is a two--fold rotation with axis perpendicular to the $C_{\infty}$ axis.
}
\begin{ruledtabular} 
\begin{tabular}{cccccc} 
state & polar vectors & axial vectors & spins (axial vectors + time reversal) & symmetry broken \\ 
\hline
2:2 collinear &$D_{\infty h}=C_{\infty}\otimes \{1,C_2',\sigma_v,I\}$& $D_{\infty h}$& $D_{\infty h} + \Theta D_{\infty h} $ & $O(3)/(O(2)\times O(1))=\mathbb{R}{\bf P}^2$ \\ 
3:1 collinear &$C_{\infty v}=C_{\infty}\times \{E,\sigma_v\}$& $C_{\infty}\times \{E,I\}$ & $C_{\infty}\times \{E,I\}+\Theta \sigma_v C_{\infty}\times \{E,I\}$ & 1 \\
2:2 canted & $C_{2 v}=\{E,C_2(z),\sigma_{xz},\sigma_{yz}\}$ & $C_{2h}=\{E,I,C_{2}(z),\sigma_{xy}\}$ & $C_{2h}+\Theta \sigma_{xz} C_{2h}$&  $C_{\infty}/C_2$\\ 
3:1 canted & $C_{1 h}=\{E,\sigma_{xz}\}$ & $S_2=\{E,I\}$ & $S_2+\Theta\sigma_{xz}S_2$&  $C_{\infty}$ \\ 
\end{tabular} 
\end{ruledtabular} 
\end{table*}

%%%%%%%%%%%%%%%%%%%%%%%%%%%%%%%%%%%%%%%%%%%%%

\section{Higher order multipoles}
\label{sec:octupoles} 

\begin{table} 
\caption{\label{tab:octupoles} Classification of rank--three tensor operators according to rotational symmetry 
about a $z$ axis defined by magnetic field.   Also indicated are the finite values of the order parameters 
in the 2:2 and 3:1 canted states.} 
\begin{ruledtabular} 
\begin{tabular}{cccc} 
order par. &tensor operators& 2:2 & 1:3  \\ 
\hline
\hline
$e^{3i\phi}$ & $\{T^{x^3-3xy^2},T^{y^3-3yx^2}\}$ & 0 & finite  \\ 
\hline
$e^{2i\phi}$ & $\{T^{z(x^2-y^2)},T^{xyz}\}$ & finite & finite   \\ 
\hline
$e^{i\phi}$ & $\{T^{x(r^2-5z^2)},T^{y(r^2-5z^2)}\}$ & 0 & finite  \\ 
\hline
1                 & $T^{z(3r^2-5z^2)}$ & finite & finite   \\ 
\end{tabular} 
\end{ruledtabular} 
\end{table} 

In this paper, we have classified states according to the {\it lowest} moment of spins which breaks spin rotational symmetry.
According to this conventional, ``common sense'' prescription, a state which lacks conventional dipolar (e.g., N\'eel) 
order, but exhibits a common plane for the canting of spins, is automatically classified as a nematic or vector--multipole phase.
While this classification scheme is unambiguous, it is not complete, and in some cases may give
the wrong answer, so far as the primary order parameter is concerned. 

This point was recently discussed at length for the {\it coplanar} ground--state manifold of the classical Heisenberg model 
on a kagome lattice, where the primary order parameter was convincingly argued to be octupolar, and {\it not} quadrupolar, 
in nature~\cite{zhitomirsky08}.   Incorrect assignment of the primary order parameter does not affect our ability to detect a 
bulk ordered phase, but can lead to false conclusions about phase transitions.   This is particularly true of two--dimensional 
systems at finite temperature, where the homotopy group associated with the order parameter determines the 
form of topological defect entering into Berezinsky--Kosterlitz--Thouless type phase transitions.  

In fact the states which we classify as ``nematic'' or ``vector--multipole'' in Sec.~\ref{deltaJzero} also posses 
higher order multipole moments which, under some circumstances, couple to the rank-two tensor order parameters
used in this paper.    We illustrate this below for the specific case of the rank--three tensor associated with
octupolar order.  

This is odd under time reversal, and has seven components 
\begin{eqnarray}
T^\alpha= \frac{1}{N} \sum_{i} T^\alpha_i
\end{eqnarray}
given by
\begin{eqnarray}
T^{x^3-3xy^2}_i  &=& (S^x_i)^3 -3 S^x_i(S^y_i)^2   \, , \\
T^{y^3-3yx^2}_i &=& (S^y_i)^3 -3 S^y_i(S^x_i)^2   \, , \\
T^{z(x^2-y^2)}_i  &=&   \sqrt{6} \left[ (S^x_i)^2 -(S^y_i)^2   \right] S^z_i \, , \\
T^{xyz}_i  &=&    2 \sqrt{6} S^x_i S^y_i S^z_i   \, , \\
T^{x(r^2-5z^2)}_i  &=& \sqrt{\frac{3}{5}} S^x_i   \left[ (S^x_i)^2+(S^y_i)^2 -4 (S^z_i)^2   \right] \, , \\
T^{y(r^2-5z^2)}_i  &=&  \sqrt{\frac{3}{5}} S^y_i   \left[ (S^x_i)^2+(S^y_i)^2 -4 (S^z_i)^2   \right] \, , \\
T^{z(3r^2-5z^2)}_i  &=&  \sqrt{\frac{2}{5}} S^z_i 
    \left[3(S^x_i)^2+3(S^y_i)^2 -2 (S^z_i)^2   \right]\, .
\label{eqn:octupoles}
\end{eqnarray}

In the absence of magnetic field, quadrupolar order can couple to (fluctuations of) octupolar order 
through terms of the form
\begin{equation}
\delta {\mathcal F} \sim \sum_{\alpha\beta\gamma\delta} Q^{\alpha\beta} T^{\alpha\gamma\delta} T^{\beta\gamma\delta}
\end{equation}
in the free energy, which respect the full $O(3)$ symmetry of the Hamiltonian, and time reversal 
invariance~\cite{lubensky02}. Therefore, a finite octupolar order parameter usually induces a quadrupolar one, while 
the opposite is not always true.  When they occur together, some care must then be taken to assign the 
correct {\it primary} order parameter.   

In finite magnetic field we again classify these octupoles according to the way in which they transform under
rotations about direction of magnetic field (the $z$ axis).    We obtain a single one--dimensional irrep and 
three two--dimensional irreps, 
\begin{eqnarray}
{\bf T}^{\perp, 3} &=&   \{T^{x^3-3xy^2} , T^{y^3-3yx^2}\}\, , \\
{\bf T}^{\perp, 2}  &=&   \{T^{z(x^2-y^2)} , T^{xyz}\}\, , \\
{\bf T}^{\perp, 1}  &=&   \{ T^{x(r^2-5z^2)} , T^{y(r^2-5z^2)} \} \, , \\
T^{\perp, 0}  &=&   T^{x^3-3xy^2} \, ,
\end{eqnarray}
which take on finite values in the different ordered states.   
These results are summarized in Table~\ref{tab:octupoles}. As the magnetic field breaks time-reversal invariance, the quadrupolar and octupolar order parameters may mix linearly in the free energy. For example, where $S^z$ is singled out by magnetic field, the new terms that enter the free energy are of the form
\begin{equation}
\delta {\mathcal F} \sim S^z \left[Q^{x^2-y^2} T^{z(x^2-y^2)} +Q^{xy} T^{xyz}\right] ,
\end{equation}
and
\begin{equation}
\delta {\mathcal F} \sim  S^z \left[Q^{xz} T^{x(r^2-5z^2)}+Q^{yz} T^{y(r^2-5z^2)}\right] ,
\end{equation}
which respect the remaining $O(2)$ rotational symmetry (more precisely, they can mix if the $S^z$ order parameter is finite, irrespectively of the presence of external magnetic field).   
Magnetic field can therefore strongly modify the symmetry of a (primary) multipolar order parameter.
For a related discussion, see Ref.~[\onlinecite{canals04}].

It is not our intention to give a definitive treatment of this complex set of coupled order parameters in 
this paper.   However we have made a preliminary study of the behavior of the rank--three and rank-four tensor 
order parameters in the present model, using the $T=0$ theory for a single tetrahedron embedded in the lattice, 
and classical Monte Carlo simulation.   We have been unable to identify any higher-order multipole which grows 
faster at a continuous transition than the rank-two tensor order parameters given in 
Section~\ref{sec:nematic_order_parameter}, and so these retain their tentative assignment as primary 
order parameters. 

%%%%%%%%%%%%%%%%%%%%%%%%%%%%%%%%%%%%%%%%%%%%%%%%%%%%%%

\end{document}